\documentclass[reqno]{amsart}

\numberwithin{equation}{section}

\usepackage{amsmath}
\usepackage{amsxtra}
\usepackage{graphicx}

\usepackage{amscd}
\usepackage{amsthm}
\usepackage{amsfonts}
\usepackage{amssymb}
\usepackage{eucal}
\usepackage{psfrag}
\usepackage{pstricks}

\newcommand{\g}{{\mathfrak{g}}}
\renewcommand{\a}{{\mathfrak{a}}}

\newcommand{\gl}{{\mathfrak{gl}}}
\newcommand{\n}{{\mathfrak{n}}}
\renewcommand{\sl}{{\mathfrak{sl}}}

\renewcommand{\a}{{\mathfrak a}}

\newcommand{\C}{{\mathbb C}}

\newcommand{\bzeta}{\boldsymbol\zeta}
\newcommand{\bmu}{\boldsymbol\mu}

\def\be{\begin{equation}}
\def\ee{\end{equation}}
\def\bea{\begin{eqnarray}}
\def\eea{\end{eqnarray}}
\def\line{\hbox to \hsize}

\def \a{{\bf a}}

\def \k{{\bf k}}

\def \n{{\bf n}}

\def \m{{\bf m}}
\def \tsh{\textstyle {\frac{1}{2}}}

\def \ket #1{{\vert #1 \rangle}}
\def \bra #1{{\langle #1\vert}}

\def\eval #1#2#3{{\langle#1\vert#2\vert#3\rangle}}

\def\1{\mbox{\bf 1}}

\newcommand{\ua}{\!\!\uparrow}
\newcommand{\da}{\!\!\downarrow}

\DeclareMathOperator{\ch}{ch}
\DeclareMathOperator{\mmod}{mod}
\newcommand{\sz}{\{z^{(i)}_j\}}
\newcommand{\zi}{z^{(i)}}
\newcommand{\zii}{z^{(i+1)}}
\newcommand{\zp}{z^{(p)}}
\newcommand{\mi}{m^{(i)}}

\newcommand{\mip}{m^{(i')}}
\newcommand{\mj}{m^{(j)}}

\newcommand{\bn}{{\bf n}}
\newcommand{\ZZ}{{\mathbb Z}}
\newcommand{\set}[2]{\{#1\ | \ #2\}}

\newcommand{\wsu}{\widehat{su}}

\newcommand{\wlam}{{\widehat\lambda}}

\newcommand{\bA}{{\bf A}}
\newcommand{\bC}{{\bf C}}
\newcommand{\bK}{{\bf K}}
\newcommand{\bN}{{\bf N}}

\newcommand{\bl}{{\bf l}}
\newcommand{\bm}{{\bf m}}
\newcommand{\bx}{{\bf x}}
\newcommand{\bz}{{\bf z}}

\newcommand{\cF}{{\mathcal F}}
\newcommand{\cK}{{\mathcal K}}
\newcommand{\cS}{{\mathcal S}}

\def\dottedhline(#1,#2)#3#4#5{%
\multiput(#1,#2)(#3,0){#4}{\circle*{#5}}
}

\newlength{\cw}
\allowdisplaybreaks[1]

\begin{document}
\title[Fusion products]
{Fusion products, Kostka polynomials, and fermionic characters of
$\widehat{su}(r+1)_k$ }

\author{Eddy Ardonne, Rinat Kedem, Michael Stone}

\begin{abstract}
Using a form factor approach, we define and compute the character of
the fusion product of rectangular representations of $\wsu(r+1)$. This
character decomposes into a sum of characters of irreducible
representations, but with $q$-dependent coefficients. We identify
these coefficients as (generalized) Kostka polynomials. Using this
result, we obtain a formula for the characters of arbitrary integrable
highest-weight representations of $\wsu(r+1)$ in terms of the
fermionic characters of the rectangular highest weight
representations.
\end{abstract}

\maketitle

\section{Introduction}

In an earlier paper \cite{usI} (which we will refer to as I) we
provided a physics application of the ideas of Feigin and Stoyanovsky (FS)
\cite{feigin-stoyanovsky}, who showed that by counting the
number of linearly independent symmetric polynomials that vanish when
$k+1$ of their variables coincide one may obtain fermionic
formul{\ae} \cite{KKMM} for the characters of integrable
representations of the affine Lie algebra $\widehat{su}(2)_k$
\cite{LP85}. We also sketched how a similar count of the
number of symmetric polynomials in two types of variables allows the
computation of the character of the vacuum representation of
$\widehat{su}(3)_k$. The reason for the restriction to the vacuum
representation of $\widehat{su}(3)_k$ was that a
na{\"\i}ve application of the FS strategy to most representations of
higher rank groups does not yield the characters. Here  we
will explain both why the na{\"\i}ve method fails, and what must be
done to correct it. To be more precise, we explain the ideas behind the
corrected
character formul{\ae} in a form accessible to physicists. We will  not
give all the technical details of the proofs. The mathematical details and the
proofs are given in \cite{usII}.

The basic idea of \cite{feigin-stoyanovsky} is most simply explained
in the language of conformal field theory: we compute matrix elements
of ladder-operator currents between the highest weight state that
defines the representation of interest and any other weight state in
the representation. These matrix elements are rational functions (in
the $\widehat{su}(2)$ case they are symmetric polynomials) in the
co-ordinates of the current operators and have certain restrictions
imposed on them by relations in the Lie algebra, and by the the
highest weight condition. By counting the number of possible functions
satisfying these constraints we are able to count the dimension of all
weight subspaces.

For $\widehat{su}(r+1)$, and for a representation whose top graded
component forms a finite-dimensional representation of $su(r+1)$
having only one non-zero Dynkin index (and hence corresponds to a
rectangular Young diagram), it is not difficult to count the
dimensions of the function space. For a general representation, it
is hard. We find, however, that by introducing a fusion-product
representation we can retain a function space whose dimensions we can
count. The price to be paid is that the fusion-product representation
is reducible. The generating function for the dimensions, although of
fermionic form, is therefore a sum of characters of irreducible
representations, moreover one with coefficients that are polynomials
in the variable $1/q$. These $q$-dependent coefficients can be
identified as being
generalized Kostka polynomials \cite{SchWar,KirShi}. The decomposition
may then be inverted to obtain the character of any desired
irreducible representation in terms of the fermionic fusion-product
characters.

Kostka polynomials and their generalizations are polynomials in $q$,
with coefficients which are non-negative integers. The classical
Kostka polynomials are known in the theory of symmetric functions
\cite{Mac} as the transition coefficients between Hall-Littlewood and
Schur polynomials.
In physics, the Kostka polynomials made their first appearance in the
study of the completeness of Bethe ansatz states in the Heisenberg
spin chain \cite{KR,KKR}. The methods introduced there, together with
the theory of crystal bases, is the basis for the subsequent studies
of \cite{KirShi,SchWar,KSS}.

In order to establish our notation, in section two we provide a brief
review of affine Lie algebras. In section three we introduce the
``principal subspace'' of Feigin and Stoyanovsky, and the function
spaces that are dual to them. In section four we show how the affine
Weyl translations allow us, in the special case of rectangular
representations, to use the character of the principal subspace to
compute the character of the full space. This is the process that in I
we called ``filling the bose sea." In section five we explain why the
na{\"\i}ve extension of the method to non-rectangular representations
fails, and exhibit the subtractions necessary to obtain
correct formul{\ae} for the characters. These formul{\ae} were
originally obtained numerically. In section six we show how these
correct formul{\ae} find their explanation in characters of reducible
fusion-product representations, and also explain the origin of the
$q$-dependent Kostka polynomial coefficients in the fusion-product
decomposition. In section seven, we make the connection between these
Kostka polynomials and the WZW conformal field theory. Finally, in
section eight, we combine all the results to obtain a character
formula for arbitrary highest-weight representations of $\wsu(r+1)_k$,
equation \eqref{chahw}. Section nine is devoted to the conclusions and
an outlook. Details concerning the characters of the
principal subspaces can be found in appendix A, while appendix B
contains details about the explicit formula for the (generalized)
Kostka polynomials.

\section{Notation}

A simple Lie algebra $\mathfrak{g}$ gives rise to the
affine Lie algebra
$\widehat{\mathfrak{g}}'= \mathfrak{g}\otimes \mathbb{C}[t,t^{-1}]\oplus
\widehat c \,\mathbb{C}$
with commutation relations
\be
[a\otimes t^m, b\otimes t^n]= [a,b]\otimes t^{n+m} +\widehat c\,
n\langle a,b\rangle \delta_{m+n,0}.
\ee
Here $\langle a,b\rangle$ denotes the Killing form,
normalized so that a long root has length $\sqrt{2}$. The element
$\widehat c$ is central, and so commutes with all
$a\in \widehat{\mathfrak{g}}$.
It is convenient to adjoin to this algebra a grading operator
$\widehat d$ that also commutes with $\widehat c$, and such that
$[\widehat d, a\otimes t^n]= n (a\otimes t^n)$. The algebra with the
adjoined operator $\widehat d$ is denoted by $\widehat{\g}$. We will
often write $a\otimes t^n\equiv a[n]$.

The affine algebra is defined for polynomials in $t,
t^{-1}$. We will later need to extend the definition to
functions which are rational functions in $t$, or more generally,
Laurent series in $t^{-1}$.
If $f(t)$ and $g(t)$ are such series, we can
write the commutation relations as
\be
[a\otimes f, b\otimes g]= [a,b]\otimes fg+
\frac{\widehat c \langle a,b\rangle}{2\pi i}\oint f'g\, dt.
\ee
Here the integral treats the otherwise formal parameter $t$ as a
complex variable taking values on a circle surrounding $t=0$. The
result of the integration, ${\rm Res}_{t=0}(f'g \,dt)$, involves only
a finite sum because of the polynomial condition on the negative
powers of $t$ in $f$, $g$.

We will use the Chevalley basis for the generators of the
simple algebra $\mathfrak{g}$, in which each simple root
$\alpha_i$ is associated with a step-up ladder operator
$e_{\alpha_i}\equiv E^{\alpha_i}$,
a step-down ladder operator
$f_{\alpha_i}\equiv E^{-\alpha_i}$,
and the co-root element of the Cartan algebra
$h_{\alpha_i}\equiv \tfrac{2 \alpha_i\cdot H}{\|\alpha_i\|^2}$.
The commutation relations of these generators are
\begin{align}
[h_{\alpha_i},h_{\alpha_j}]&=0 \ , &
[h_{\alpha_i},e_{\alpha_j}]&=(\bC_r)_{ji}
e_{\alpha_j}\ , \nonumber \\
[h_{\alpha_i},f_{\alpha_j}]&=-(\bC_r)_{ji}
f_{\alpha_j}\ , &
[e_{\alpha_i},f_{\alpha_j}]&= \delta_{ij}
h_{\alpha_i}\ .
\end{align}
The remaining generators are obtained by repeated
commutators of these, subject to the {\em Serre relations}
\be
\label{serrel}
[{\rm ad}(e_{\alpha_i})]^{1-(\bC_r)_{ji}} e_{\alpha_j}=0,\quad
[{\rm ad}(f_{\alpha_i})]^{1-(\bC_r)_{ji}} f_{\alpha_j}=0.
 \ee
In these expressions $(\bC_r)_{ij}$ denotes the elements of the Cartan
matrix of $\mathfrak{g}$:
\be
(\bC_r)_{ij}\stackrel{\rm def}{=}
\frac{2 (\alpha_i\cdot \alpha_j)}
{\phantom 2\|\alpha_j\|^2}.
\ee
In the case
$\mathfrak{g}= su(r+1)$ the elements of the Cartan matrix are
given by $(\bC_r)_{ij} = 2 \delta_{i,j}- \delta_{|i-j|,1}$.
We will also have cause to use the
inverse Cartan matrix whose elements (for $\mathfrak{g}=
su(r+1)$) are
$(\bC^{-1}_r)_{ij} = \min(i,j) - \tfrac{i j}{r+1}$.

Our interest is in {\it integrable\/} representations of
$\widehat{\mathfrak{g}}$, and in particular of
$\widehat{su}(r+1)$. An integrable representation is one
that, under restriction to any subgroup of
$\widehat{\mathfrak{g}}$ isomorphic to $su(2)$, decomposes into a set of
finite-dimensional representations of this $su(2)$. From the work
of Kac \cite{kac} it is known that in any
irreducible integrable representation the generator
$\widehat c$ will act as a positive integer multiple of the identity,
$ \widehat c \mapsto k \mathbb{I}$,
and we will follow the physics convention of
appending the integer $k$, the {\it level\/} of the representation,
to the name of the group, and so
write $\widehat{\mathfrak{g}}_k$. These integrable
representations are all highest-weight representations
whose highest weight vector is annihilated by
$e_{\alpha_i}[n]$, $n\ge 0$ and by
$h_{\alpha_i}[n]$, $f_{\alpha_i}[n]$ with $n>0$. The top d-graded
subspaces (where $\widehat d$ is taken to act as zero) form
finite-dimensional 
representations of the simple algebra $\mathfrak{g}$, but
not all representations of $\mathfrak{g}$ can form top
components of integrable representations of
$\widehat{\mathfrak{g}}_k$. In the case of
$\widehat{su}(r+1)_k$ the restriction on the
representations of $su(r+1)$ that can appear as top components
is that the number of columns in the Young diagram
labeling the representation must be $\le k$.

We wish to obtain the dimensions ${\rm mult}(\widehat{\mu})$
of the weight spaces
$\widehat{\mu}\equiv (\mu;k;d)$
in an irreducible integrable level-$k$ representation
$H_{\widehat{\lambda}}$ of
$\widehat{\mathfrak{g}}_k$, whose highest weight is
$\widehat{\lambda}=(\lambda;k;0)$.
Here $\lambda$ and $\mu$ denote a
highest weight and an arbitrary weight, respectively, of
the associated finite-dimensional simple algebra $\mathfrak{g}$, and
$d$, a non-positive integer, is the eigenvalue of the
grading operator $\widehat d$. The character of the
representation is then defined to be
\be
{\rm ch\,}H_{\widehat{\lambda}} (q,\bx) =
\sum_{\widehat{\mu}} {\rm mult}(\widehat{\mu})
\,x_1^{\mu_1} x_2^{\mu_2}\cdots x_r^{\mu_r}\,q^{-d}.
\ee
Here $\mu_1,\ldots,\mu_r$ are the
components of the weight $\mu$ in the basis of
the fundamental weights of the finite-dimensional algebra
$\omega_i = \sum_j (\bC_r^{-1})_{ji} \alpha_j$,
{\it i.e.\/}
$\mu=\sum_{i=1}^{r} \mu_i \omega_i$, and
$\bx=(x_1,\ldots,x_r)$. Because the eigenvalues $d$ are zero or
negative, the character is a formal series in positive powers of $q$.

In this paper, we present fermionic formulas for the characters of all
highest weight integrable representations of $\widehat{su}(r+1)$. These
formulas are inspired by the construction of Feigin and Stoyanovsky,
which we present below.

\section{The principal subspace}
\label{sec_pss}

In this section we briefly review the strategy of Feigin and
Stoyanovsky \cite{feigin-stoyanovsky}, as we presented it in I.
The reader should refer to I for more details.

Let $\ket{\widehat{\lambda}}$ be the
highest-weight state of the irreducible representation $H_\wlam$.
Recall that
$e_{\alpha_i}$ and $f_{\alpha_i}$ are the step-up and
step-down ladder operators corresponding to simple root
$\alpha_i$ in the finite-dimensional algebra
$\mathfrak{g}$, and $e_{\alpha_i}[n]$ and
$f_{\alpha_i}[n]$ are their $\widehat{\mathfrak{g}}$ relatives.
Since $H_\wlam$ is irreducible,
by acting on $\ket{\widehat{\lambda}}$ with products of
$e_{\alpha_i}[n]$, $n<0$, and
$f_{\alpha_i}[n]$, $n\le 0$, we generate
the entire space $H_{\widehat{\lambda}}$. If
we restrict ourselves to products containing only the
$f_{\alpha_i}[n]$, $n\le 0$, we obtain what
is known as the {\it principal subspace\/} $W_{\widehat{\lambda}}$.

Note that in I we used a commuting set of operators applied to the
highest weight state to generate a principal subspace (in the case of
$su(2)$ and $su(3)$). This subspace had the advantage for physics
applications that the resulting correlation functions were
polynomials, rather than rational functions. For the purposes of this
paper, however, it is necessary to use the Feigin-Stoyanovsky
subspace, in order to obtain the simple closed-form formul\ae\ for the
principal subspace multiplicities for arbitrary representations that
we report here.

Consider, for example, $su(2)$ where
there is only one simple root $\alpha_1$ and so the root label is
redundant. Here, since each application of an $f[-n]$ moves us one
step to the left and $n$ steps down in the weight diagram, we cannot
reach any weight to the right of the column lying below
$\lambda$. (See figure \ref{su2k2vac} where the upper part of the
weight diagram of the vacuum representation of $\wsu(2)_2$ is
displayed.) Furthermore, we can reach weights close to this column
{\it via\/} fewer distinct products $f[-n_1]f[-n_2]\ldots f[-n_m]$
than the dimension of the weight space. Because of this paucity of
paths, even if all the $f[-n_1]f[-n_2]\ldots
f[-n_m]\ket{\widehat{\lambda}}$ were linearly independent, we would be
able to obtain only a subspace in each weight
space. If, however, we look at weights in columns far to the left in
the weight diagram, the number of distinct paths leading to a given
weight grows rapidly, whilst the dimensions of the weight spaces near
the head of any such column remain small. It is plausible, and indeed
true, that we can obtain all states in such a weight space by applying
suitable products of $f[-n]$'s to $\ket{\widehat{\lambda}}$. Thus, by
counting the number of linearly independent paths, we can obtain the
multiplicity of these distant weights. Because the weight diagram is
invariant under the action of affine Weyl translations,
as we explain in section \ref{sec_awt}, this
information provides the dimension of {\it all} the weight spaces, and
hence the complete character.

We could in principal use Lie algebra relations to
determine the number of independent paths between the
weights. It is easier, however, to obtain the dimensions
of the weights $\widehat{\mu}$ appearing in
the principal subspace by
counting the number of linearly independent functions $F$
that arise as matrix elements
\be
\label{matelem}
F(\{z^{(1)}\},\{z^{(2)}\},\ldots)=
\eval{v}
{R\{f_{\alpha_1} (z_{1}^{(1)})\ldots f_{\alpha_1}(z_{m^{(1)}}^{(1)})
f_{\alpha_2}(z_{1}^{(2)})\ldots f_{\alpha_2}(z_{m^{(2)}}^{(2)})
\ldots\}}{\widehat{\lambda}},
\ee
where $\ket{v}$ is an element of a weight space $\widehat{\mu}$.

In addition, $f_{\alpha_i}(z)$ denotes the current operator
\be
\label{curop}
f_{\alpha_i}(z)= \sum_{n=-\infty}^{\infty} f_{\alpha_i}[n] z^{-n-1},
\ee
and the $R$ denotes ``radial ordering'', meaning that operators $f(z)$
with larger $|z|$ are placed to the left of those with smaller
$|z|$'s%
\footnote{Strictly speaking, we regard $z_i^{(\alpha)}$'s as
formal variables and consider matrix elements of all possible orderings
of the roots $\alpha_i$. We then count the resulting number of linearly
independent functions in the formal variables $z_i^{(\alpha)}$.}.
By duality, the number of linearly independent functions $F$
with $m^{(i)}$ currents $f_{\alpha_i} (z)$ and fixed total degree $N$
is equal to the dimension, in the principal subspace $W_\wlam$
of the weight
\be
\label{pssweight}
\widehat{\mu}=
\left(\lambda-\sum_i m^{(i)}\alpha_i;k;-N -\sum_i m^{(i)}\right).
\ee
Thus, the numbers $\mi$ correspond to the number of roots $\alpha_i$
which need to be subtracted from $\lambda$ to obtain $\mu$, hence, in
terms of the fundamental weights, we have
$\mu = \sum_{i=1}^{r} (l_i - \sum_j(\bC_r)_{ij} m^{(j)}) \omega_i$,
where the $l_i$ are the components of $\lambda$, {\it i.e.\/}
$\lambda= \sum_i l_i\omega_i$.

The linear dependencies between products of $f_{\alpha_i}[n]$
due to relations in the Lie algebra, the integrability, and
the properties of the highest weight vector
$\ket{\widehat{\lambda}}$ translate, through duality, into
conditions satisfied by the function $F$. For example,
operators $f_{\alpha_i}(z)$ corresponding to the same
simple root commute with each other. The function $F$ is
therefore a symmetric function in the $m^{(i)}$
variables $z_{j}^{(i)}$ for fixed $i$. It must also possess
certain poles and zeros whose exact form we will specify
below. Once we know all these properties, we can set out to
count the number of independent functions. It is not,
however, easy to be sure that we have obtained a complete
set of constraints on $F$. If we have have missed some, we
will over-count the multiplicities.

In the case of $\widehat{su}(r+1)_k$ representations with only
one non-zero Dynkin index, {\it i.e.\/} $\lambda = l \omega_p$,
it is not too hard to find all the constraints on
functions $F(\sz)$.
We already observed that $F(\sz)$ is symmetric under the
exchange of the variables $\zi_j \leftrightarrow \zi_{j'}$.
Next we observe that the commutator
\be
[f_{\alpha_{i}}, f_{\alpha_{i+1}}]=
f_{\alpha_{i}+\alpha_{i+1}}
\ee
implies the
operator product
\be
f_{\alpha_{i}} (z) f_{\alpha_{i+1}} (w) =\frac{
f_{\alpha_i+\alpha_{i+1}} (w)}{(z-w) } +\hbox{regular terms},
\ee
and this in turn
implies that $F(\sz)$ can have a pole of at most order one
when the coordinates of two currents corresponding to adjacent
simple roots coincide, $\zi_{j}=\zii_{j'}$. (There is no
pole for non-adjacent indices because
$\alpha_i+\alpha_{j}$ is not a root unless $j=i\pm1$.)
We will refer to the index $i$ as the {\em color} of the
variables. In the representation $\lambda = l \omega_p$,
and for $l>0$, we have that $f_{\alpha_p}[0]
\ket{\wlam}\ne0 $, and since $f_{\alpha_p}[0]$ comes with
coefficient $z^{-1}$, the function $F(\sz)$ may have a
pole of order one at $z^{(p)}_j = 0$. (Note that 
$f_{\alpha_i}[0] \ket{\wlam} = 0$ for $i\neq p$.)
These are the only
possible poles, and we make them explicit by writing
\begin{equation}
\label{ratfunc}
F(\sz) = \frac{f(\sz)}
{\prod_{j}(z^{(p)}_j) \prod_{i=1}^{r-1}\prod_{j,j'} (\zi_j-\zii_{j'})} \ ,
\end{equation}
where $f(\sz)$ is a now a {\it polynomial\/} symmetric
under the exchange of variables of the same color: $\zi_j
\leftrightarrow \zi_{j'}$. Because of relations in the
algebra and properties of the representation $f(\sz)$ is not an
arbitrary symmetric polynomial, but must possess certain zeros. These
we now describe:

\begin{itemize}

\item[i)]
The integrability
condition requires that $[f_{\alpha_i}(z)]^{k+1}$ annihilates
{\em any} vector in the representation. This tells us that
$f(\sz) = 0$ when $\zi_1=\zi_2=\cdots=\zi_{k+1}$, for any
color $i$.
\item[ii)] The integrability properties of the top component of the
representation with highest weight $\lambda =l \omega_p$ tell us that
$f_{\alpha_p}^{l+1}[0]\ket{\wlam}=0$. The function $f(\sz)$ must therefore
have a zero when  $l+1$ of the $\zp_{j}$ become zero. 
\item[iii)]
The Serre relations \eqref{serrel} are
\be
[f_{\alpha_i},[f_{\alpha_i},f_{\alpha_{i+1}}]]=
[f_{\alpha_{i+1}},[f_{\alpha_{i+1}},f_{\alpha_i}]]=0,
\ee
and indicate that $F$ should have no pole if two currents of
color $i$ are  made to coincide
with a current of adjacent color. This requires
that $f(\sz)$ have a pole-canceling zero when
$\zi_1=\zi_2=\zii_1$ or when $\zi_1=\zii_1=\zii_2$, for
$i=1,\ldots,r-1$.
\end{itemize}

To summarize: we find that the polynomial $f(\sz)$ is symmetric in the
variables corresponding to the same color, and vanishes
when any of the following conditions holds
\begin{align}
\zi_1&=\cdots=\zi_{k+1} \ , \quad \forall i \label{intc}\\
\zi_1&=\zi_2=\zii_1 \ , \ \zi_1=\zii_1=\zii_2 \ , \quad
i = 1,\ldots, r-1 \label{serrec}\\
\zp_1&= \cdots = \zp_{l+1} = 0 \label{topc} \ .
\end{align}
We will denote the space of rational functions $F(\sz)$ \eqref{ratfunc}
such that $f(\sz)$ satisfies \eqref{intc}, \eqref{serrec} and
\eqref{topc} by ${\mathcal F}_{l \omega_p;k}$.
We now define the character of this space to be 
\begin{equation}
\label{funchar}
\ch {\mathcal F}_{l \omega_p;k} (q,\bx) \stackrel{\rm def}{=}
x_p^{l}
\sum_{\{F(z)\}} q^{\deg(F) + \sum_{i=1}^{r} \mi}
\bigl(\prod_{i=1}^{r} x_i^{-\sum_j (\bC_r)_{ji} \mj}\bigr) \ ,
\end{equation}
where the sum is over all the functions $F(\sz)$ in the space
${\mathcal F}_{l \omega_p;k}$.
The powers of $q$ and $x_i$ are motivated by the 
form of the weights \eqref{pssweight}, which we rewrite here
in terms of the fundamental weights $\omega_i$
\begin{equation}
\widehat{\mu}=
\left(\lambda-\sum_{i,j} m^{(j)} (\bC_r)_{ji} \omega_i 
;k;-N -\sum_i m^{(i)}\right).
\end{equation}

As a reminder, $\mi$ is the number of variables of color $(i)$, which
corresponds to the number of roots $\alpha_i$ subtracted from the
highest weight $\lambda=l \omega_p$.
Note that the exponent of $q$ is not just the total
degree of $F(z)$, but is shifted by $\sum_i \mi$ (compare with
\eqref{pssweight}). The reason is that
the currents $f_{\alpha_i} (z)$ are defined by eq. \eqref{curop},
in which the power of $z$ is shifted by the scaling dimension.

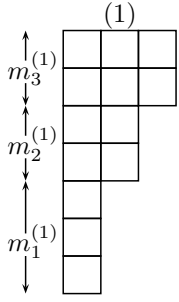
\begin{figure}[ht]
\begin{center}
\psset{unit=1mm,linewidth=.2mm,dimen=middle}
\begin{pspicture}(-2,0)(20,40)
\multips(5,0)(0,5){7}{\psframe(0,0)(5,5)}
\multips(10,15)(0,5){4}{\psframe(0,0)(5,5)}
\multips(15,25)(0,5){2}{\psframe(0,0)(5,5)}
\rput(1,7){$m^{(1)}_1$}
\rput(1,20){$m^{(1)}_2$}
\rput(1,30){$m^{(1)}_3$}
\psset{arrows=->}
\psline(0,5)(0,0)
\psline(0,8)(0,15)
\psline(0,18)(0,15)
\psline(0,21)(0,25)
\psline(0,28)(0,25)
\psline(0,31)(0,35)
\rput(12.5,37){$(1)$}
\end{pspicture}
\end{center}
\caption{A partition of $m^{(1)}=13$.}
\label{fig_partition}
\end{figure}

Counting the dimensions of these function spaces is an intricate but
tractable problem. We will explain the structure of this character in
the appendix \ref{app_pss}, and refer to \cite{usII} for details and
the proof. The result is the function-space character
\begin{equation}
\label{psschar}
\ch {\mathcal F}_{l \omega_p;k} (q,\bx) =
x_p^l
\sum_{\substack{i=1,\ldots,r\\a=1,\ldots,k\\\mi_a\geq 0}}
\Bigl(\prod_{i=1}^r x_i^{- \sum_j (\bC_r)_{ji}\mj}\Bigr)
\frac{q^{\frac{1}{2} m^{(i)}_a (\bC_r)_{i,j} \bA_{a,b} m^{(j)}_{b} -
\bA_{a,l} m^{(p)}_a}}
{\prod_{i=1}^r \prod_{a=1}^k (q)_{\mi_a}} \ ,
\end{equation}
where it is understood that repeated indices are summed
over. Here various symbols need to be defined: i) the
matrix $\bA$ has the entries $\bA_{a,b} = \min (a,b)$; ii) for any
integer $m>0$, we define
$(q)_m = \prod_{i=1}^{m} (1-q^i)$ and $(q)_0 = 1$; iii)
The sum over integers $\mi_a$ is to be understood as a sum
over partitions of $\mi$, where $\mi_a$ denotes the number of rows of
length $a$ in partition $(i)$. That is
$\sum_{a=1}^k a \mi_a = m^{(i)}$, see figure \ref{fig_partition} for
an example.

The space of functions $\cF_{l\omega_p;k}$ is
dual to the principal subspace $W_{l\omega_p;k}$. It follows that the
character \eqref{psschar} is the character of the principal subspace
\begin{equation}
\ch W_{l\omega_p;k} (q,\bx) = \ch \cF_{l\omega_p;k} (q,\bx) \ .
\end{equation}
In the next section, we will explain how we can use the affine Weyl
translations to obtain characters for the full representation from the
characters of the principal subspaces.

We would like to note that it is straightforward to generalize the
function space $\cF_{l\omega_p;k}$, by changing the constraint
\eqref{topc} to $\zi_1= \cdots = \zi_{l_i+1} = 0$ for
$i=1,\ldots,r$ and taking the product over all $p$ in the denominator
of \eqref{ratfunc}.
We will denote this function space by $\cF_{\bl;k}$,
where $\bl=(l_1,\ldots,l_r)^T$. The character of this function space is
\begin{equation}
\label{genlfunc}
\ch {\mathcal F}_{\bl;k} (q,\bx) =
\prod_{i=1}^{r} x_i^{l_i}
\sum_{\substack{i=1,\ldots,r\\a=1,\ldots,k\\\mi_a\geq 0}}
\Bigl(\prod_{i=1}^r x_i^{- \sum_j (\bC_r)_{ji}\mj}\Bigr)
\frac{q^{\frac{1}{2} m^{(i)}_a (\bC_r)_{i,j} \bA_{a,b} m^{(j)}_{b} -
\bA_{a,l_i} m^{(i)}_a}}
{\prod_{i=1}^r \prod_{a=1}^k (q)_{\mi_a}} \ .
\end{equation}
We would like to stress that this character not the character of the
principal subspace $W_{\lambda;k}$, where $\lambda = \sum_i l_i
\omega_i$. However, it can be interpreted as a character of a somewhat
larger space, as we will see below.

\section{Affine Weyl translations}
\label{sec_awt}

In this section, we will exploit the symmetry of the representation
spaces under the affine Weyl group. Elements of the affine Weyl group
map weights of a highest weight representation to other weights in the
same representation, in such a way that the weight-space
multiplicities are preserved.

Elements of the affine Weyl group can be thought of as a
product of a finite Weyl reflection and an affine Weyl
translation. We will focus on the abelian subgroup generated by the
affine Weyl translations, because
they will enable us to obtain the character of the full integrable
representation from the character of the principal subspace.

The affine Weyl translation $T^{N_i}_{\alpha_i}$ acts on a
weight $\hat\lambda =(\lambda;k;d)$, where $\lambda=\sum_i l_i \omega_i$,
by `translating' it to $\lambda+N_i \alpha_i$ (no summation implied)
and shifting the value of $d$
(see, for instance, \cite{kac}, equation (6.5.2))
\begin{equation}
T^{N_i}_{\alpha_i} (\lambda;k;d) =
(\lambda + k N_i \alpha_i ; k ; d - N_i l_i - N_i^2 k) \ .
\end{equation}
More generally, we have
\begin{equation} \label{awt}
\prod_{i=1}^{r} T^{N_i}_{\alpha_i} (\lambda;k;d) =
(\lambda + k \sum_{i=1}^{r} N_i \alpha_i ; k ;
d - \sum_{i=1}^{r} N_i l_i - \frac{k}{2} \bN^T \cdot \bC_r \cdot \bN)
\ ,
\end{equation}
where $\bN = (N_1,\ldots,N_r)^T$, and $\bC_r$ is the Cartan matrix of
$su(r+1)$.

\begin{figure}
\psset{unit=1mm}
\begin{pspicture}(-12,-5)(65,70)
\psset{dimen=middle}
\psline(-6,-5)(-6,60)(65,60)
\newgray{mygray}{.8}
\psset{fillstyle=solid,fillcolor=lightgray}
\multirput(0,0)(10,0){7}{\pscircle(0,0){2.5}}
\multirput(0,10)(10,0){7}{\pscircle(0,0){2.5}}
\multirput(10,20)(10,0){5}{\pscircle(0,0){2.5}}
\multirput(10,30)(10,0){5}{\pscircle(0,0){2.5}}
\multirput(20,40)(10,0){3}{\pscircle(0,0){2.5}}
\multirput(20,50)(10,0){3}{\pscircle(0,0){2.5}}
\pscircle(10,40){3.5}
\pscircle(50,40){3.5}
\pscircle(30,60){3.5}
\psset{fillcolor=black}
\multips(-6,0)(0,10){6}{\pscircle(0,0){.5}}
\multips(0,60)(10,0){3}{\pscircle(0,0){.5}}
\multips(40,60)(10,0){3}{\pscircle(0,0){.5}}
\psset{fillstyle=none}
\rput(0,0){$2$}
\rput(10,0){$10$}
\rput(20,0){$21$}
\rput(30,0){$28$}
\rput(40,0){$21$}
\rput(50,0){$10$}
\rput(60,0){$2$}
\rput(0,10){$1$}
\rput(10,10){$5$}
\rput(20,10){$13$}
\rput(30,10){$16$}
\rput(40,10){$13$}
\rput(50,10){$5$}
\rput(60,10){$1$}
\rput(10,20){$3$}
\rput(20,20){$7$}
\rput(30,20){$10$}
\rput(40,20){$7$}
\rput(50,20){$3$}
\rput(10,30){$1$}
\rput(20,30){$4$}
\rput(30,30){$5$}
\rput(40,30){$4$}
\rput(50,30){$1$}
\rput(10,40){$v_{-1}$}
\rput(20,40){$2$}
\rput(30,40){$3$}
\rput(40,40){$2$}
\rput(50,40){$v_{1}$}
\rput(20,50){$1$}
\rput(30,50){$1$}
\rput(40,50){$1$}
\rput(30,60){$v_{0}$}
\rput(-10,0){$-6$}
\rput(-10,10){$-5$}
\rput(-10,20){$-4$}
\rput(-10,30){$-3$}
\rput(-10,40){$-2$}
\rput(-10,50){$-1$}
\rput(-12,60){$d=0$}
\rput(0,68){$-6$}
\rput(10,68){$-4$}
\rput(20,68){$-2$}
\rput(30,68){$0$}
\rput(40,68){$2$}
\rput(50,68){$4$}
\rput(60,68){$6$}
\rput(-7,68){$\lambda =$}
\psset{arrowinset=.6,arrowsize=1mm 3}
\psline{<-}(32.47,57.52)(38.23,51.77)
\psline{<-}(41.77,48.23)(47.53,42.47)
\psline{<-}(51.11,36.68)(59.21,12.37)
\psline{<-}(60.79,7.63)(65,-5)
\psline{<-}(-5,-5)(-0.79,7.63)
\psline{<-}(0.79,12.37)(8.89,36.68)
\psline{<-}(12.47,42.47)(18.23,48.23)
\psline{<-}(21.77,51.77)(27.53,57.53)
\rput(39,56){$f[1]$}
\rput(48,47){$f[1]$}
\rput(59,25){$f[3]$}
\rput(66,4){$f[3]$}
\rput(0,25){$f [-3]$}
\rput(11,48){$f[-1]$}
\rput(20,57){$f[-1]$}
\end{pspicture}
\caption{The top part of the weight diagram of the vacuum representation of
$\widehat{su}(2)_2$. The numbers denoted the dimension of the
corresponding weight space.}
\label{su2k2vac}
\end{figure}
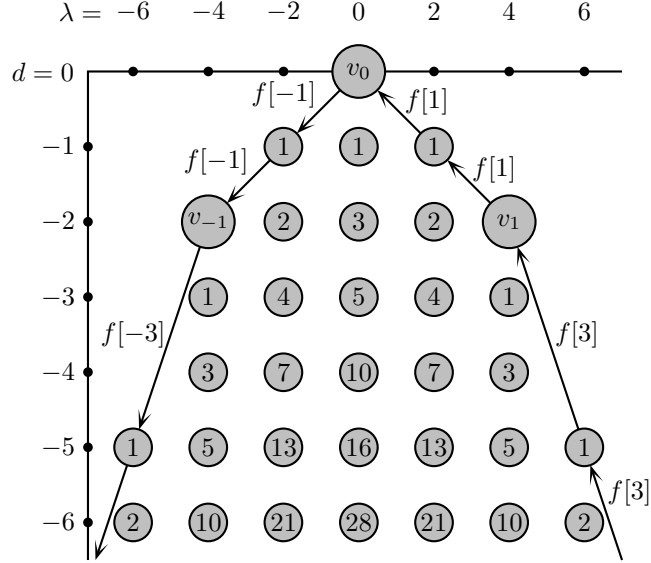

We can use these affine Weyl translations to obtain the characters of
the full representation. To illustrate how this works, we will use an
explicit example, namely the vacuum representation of $\wsu(2)_2$. The
top part of the weight diagram of this representation is given in figure
\ref{su2k2vac}.

From this figure, we see that we can obtain the whole representation
by acting
with an arbitrary combination of the operators $e[i]$ and $f[i]$, with
$i<0$ on the highest weight state
$\ket{\hat\lambda}\equiv\ket{v_0}$ (we dropped the subscript
$\alpha_1$). Under the affine Weyl
translation $T^1$, this weight is mapped to the weight $\ket{v_1}$. We can
also generate the whole representation by acting with operators $e[i]$ and
$f[j]$ on this state. In this case, we need to act with an arbitrary
combination of operators taken from the sets
$\set{e[i]}{i<-2}$ and $\set{f[j]}{j<2}$. In general, the whole
representation can be obtained by acting on the state
$\ket{v_{N}}=T^N \ket{v_0}$ with the operators
$\set{e[i]}{i<-2 N}$ and $\set{f[j]}{j<2N}$. We can now take the limit
$N\rightarrow\infty$, and obtain that, we can
generate the whole representation by acting on $\ket{v_{\infty}}$ with
only the step down operators $\set{f[j]}{j\in\ZZ}$, see 
\cite{Pri,FJLMM} for the proof of this statement.

Let us come back to the principal subspace. As a reminder, the
principal subspace $W_\wlam$ is the space generated by acting with
the operators $f[j]$, with $j<0$ on $\ket{v_0}$ (again, we are focusing on
the vacuum representation of $\widehat{su}(2)_2$). We can define a
sequence of subspaces, by acting with the affine Weyl translation,
{\it i.e.\/} $W^{(N)} = T^{N} W_\wlam$. The subspace $W^{(N)}$
is obtained by
acting with the operators $\set{f[j]}{j< - 2N}$ on the state
$\ket{v_{N}}$. So, in the limit $N\rightarrow\infty$, we find that
$W^{(\infty)}$ is obtained by acting with $\set{f[j]}{j\in\ZZ}$, on
the state $\ket{v_{\infty}}$. Comparing this subspace with the description of
the full representation of the previous paragraph, we find that they
are in fact the same.

Using this result, we can obtain the characters of the integrable
representation by acting with the affine Weyl translation $T^{N}$ on the
character of the principal subspace, and taking the limit
$N\rightarrow\infty$.

In our paper \cite{usI}, we showed that the effect of acting with the
affine Weyl translation and taking the limit of $N\rightarrow\infty$
only alters the character of the principal subspaces of $\wsu(2)_k$
in the following two ways.
First of all, the summation over the variable $m^{(1)}_k \geq 0$ is
extended to the negative integers. Secondly, the factor
$\tfrac{1}{(q)_{m^{(1)}_k}}$ is replaced by $\tfrac{1}{(q)_\infty}$.
In physical terms (see \cite{usI}), this corresponds to `filling the
Bose sea', and considering the excitations on top of a `large
droplet'. In \cite{usII}, we showed that this procedure also works
in the case $\wsu(r+1)_k$. Using this result, we obtain the
characters for rectangular highest-weight representations
\begin{equation}
\label{Hcharm}
\ch H_{l \omega_p;k} (q,\bx)=
\frac{1}{(q)_\infty^r}
x_p^l
\sum_{\substack{\mi_k \in \ZZ\\\mi_{a<k} \in \ZZ_{\geq0}\\i=1,\ldots,r}}
\Bigl(\prod_{i=1}^r x_i^{ -\sum_j (\bC_r)_{ji} \mj  }\Bigr)
\frac{q^{\frac{1}{2} \mi_a (\bC_r)_{i,j} \bA_{a,b} \mj_{b} -
\bA_{a,l} m^{(p)}_a}}
{\prod_{i=1}^r \prod_{a=1}^{k-1} (q)_{\mi_a}} \ .
\end{equation}

The character \eqref{Hcharm} can be
written in a form which allows all $l_i$ to be non-zero, namely, in
the same way as the character $\ch \cF_{\bl;k}$, equation \eqref{genlfunc}
\begin{equation}
\label{finfchar}
\ch \cF^\infty_{\bl;k} (q,\bx) \stackrel{\rm def}{=}
\frac{1}{(q)_\infty^r}
\bigl(\prod_{i=1}^{r} x_i^{l_i}\bigr)
\sum_{\substack{\mi_k \in \ZZ\\\mi_{a<k} \in \ZZ_{\geq0}\\i=1,\ldots,r}}
\Bigl(\prod_{i=1}^r x_i^{-\sum_j (\bC_r)_{ji} \mj}\Bigr)
\frac{q^{\frac{1}{2} \mi_a (\bC_r)_{i,j} \bA_{a,b} \mj_{b} -
\bA_{a,l_i} \mi_a}}
{\prod_{i=1}^r \prod_{a=1}^{k-1} (q)_{\mi_a}} \ .
\end{equation}
We cannot however interpret this character as the character of
the highest-weight representation $H_\wlam$, where the finite part of
$\hat{\lambda}$ is given by $\lambda = \sum_i l_i \omega_i$. This is
because we did not take all the constraints into account for arbitrary
representations. For example, for general highest weight, our
restrictions on the space do not ensure that
$f_\alpha [1] \ket{\wlam} = 0$ whenever
$\alpha$ is not a simple positive root. However, as
we will show in the next section, the characters
$\ch \cF^\infty_{\bl;k}$ are an essential ingredient of the characters
$\ch H_{\wlam;k}$ for arbitrary highest weights.

\section{Arbitrary highest-weight representations}
\label{sec_ahwm}

As mentioned above, we do not expect the function-space characters
$\ch \cF^{\infty}_{\bl;k}$ defined in eq. \eqref{finfchar} to be the
characters of arbitrary highest-weight representations $H_{\wlam}$.
This is because we have not yet imposed all the conditions
on the function spaces. We have not imposed
the remaining conditions, because the complexity of the resulting
space makes counting its dimensions intractable. 

We anticipate, therefore, that the function-space character
\eqref{finfchar} over-counts the dimensions of the weight spaces
whenever $\lambda$ is not a rectangular representation. Nevertheless
it is reasonable to conjecture some sort of relation between $\ch
H_{\wlam}$ and $\ch \cF^{\infty}_{\bl;k}$.

To seek such a relation we used
${Mathematica}^{\mbox{\textregistered}}$ to compare the
multiplicities given by $\ch {\mathcal F}^{\infty}_{\bl;k}$ with those
in the representation $H_{\wlam}$, obtained by using the affine version
of Freudenthal's recursion formula (see for instance,
\cite{FMS}, page 578). We found that whenever more than one
Dynkin index is non-zero, the character $\ch \cF^\infty_{\bl;k}$ does
indeed over-count, and subtractions are necessary.  It turns out that
these subtractions can be written as a sum over $\ch {\mathcal
F}^{\infty}_{\bl';k}$, where the coefficients are polynomials in
$q^{-1}$.

To illustrate this we give a table expressing the
characters of integrable level-$4$ representations of
$\widehat{su}(4)$ in terms of the characters
$\ch \cF_{\bl';k}^{\infty}$.
\begin{align}
\ch H_{1,1,0;4} &= \ch {\mathcal F}^\infty_{1,1,0;4} - \frac{1}{q}
\ch {\mathcal F}^\infty_{0,0,1;4} \nonumber \\
\ch H_{2,1,0;4} &= \ch {\mathcal F}^\infty_{2,1,0;4} -\frac{1}{q}
\ch {\mathcal F}^\infty_{1,0,1;4} +
\frac{1}{q^2} \ch {\mathcal F}^\infty_{0,0,0;4} \nonumber\\
\ch H_{1,2,0;4} &= \ch {\mathcal F}^\infty_{1,2,0;4} -\frac{1}{q}
\ch {\mathcal F}^\infty_{0,1,1;4} +
\frac{1}{q^2} \ch {\mathcal F}^\infty_{1,0,0;4} \nonumber\\
\ch H_{1,1,1;4} &= \ch {\mathcal F}^\infty_{1,1,1;4} -
\left(\frac{1}{q} + \frac{1}{q^2}\right) \ch {\mathcal
F}^\infty_{0,1,0;4} - \frac{1}{q} \ch {\mathcal F}^\infty_{2,0,0;4}
- \frac{1}{q} \ch {\mathcal F}^\infty_{0,0,2;4} \nonumber\\
\ch H_{3,1,0;4} &= \ch {\mathcal F}^\infty_{3,1,0;4} - \frac{1}{q}
\ch {\mathcal F}^\infty_{2,0,1;4} +
\frac{1}{q^2} \ch {\mathcal F}^\infty_{1,0,0;4} \nonumber\\
\ch H_{2,2,0;4} &= \ch {\mathcal F}^\infty_{2,2,0;4} -
\frac{1}{q} \ch {\mathcal F}^\infty_{1,1,1;4} +
\left(\frac{1}{q^2} + \frac{1}{q^3}\right)\ch {\mathcal F}^\infty_{0,1,0;4} +
\frac{1}{q^2} \ch {\mathcal F}^\infty_{2,0,0;4} \nonumber\\
\ch H_{2,1,1;4} &= \ch {\mathcal F}^\infty_{2,1,1;4} -
\left(\frac{1}{q}+\frac{1}{q^2}\right)\ch {\mathcal F}^\infty_{1,1,0;4} -
\frac{1}{q} \ch {\mathcal F}^\infty_{3,0,0;4} -
\frac{1}{q} \ch {\mathcal F}^\infty_{1,0,2;4} + \nonumber \\* & \quad
\left(\frac{1}{q^2}+\frac{1}{q^3}\right) \ch {\mathcal F}^\infty_{0,0,1;4}
\nonumber\\
\ch H_{1,3,0;4} &= \ch {\mathcal F}^\infty_{1,3,0;4} -
\frac{1}{q} \ch {\mathcal F}^\infty_{0,2,1;4} +
\frac{1}{q^2} \ch {\mathcal F}^\infty_{1,1,0;4} -
\frac{1}{q^3} \ch {\mathcal F}^\infty_{0,0,1;4}
\nonumber \\
\ch H_{1,2,1;4} &= \ch {\mathcal F}^\infty_{1,2,1;4} -
\left(\frac{1}{q}+\frac{1}{q^2}\right) \ch {\mathcal F}^\infty_{0,2,0;4} -
\frac{1}{q} \ch {\mathcal F}^\infty_{2,1,0;4} -
\frac{1}{q} \ch {\mathcal F}^\infty_{0,1,2;4} +
\frac{1}{q^2} \ch {\mathcal F}^\infty_{1,0,1;4} -
\nonumber \\* & \quad
\frac{1}{q^3} \ch {\mathcal F}^\infty_{0,0,0;4} \ .
\end{align}
In addition, we found evidence for the relation
\begin{equation}
\ch H_{l_1,0,l_3;k} = \ch \cF^\infty_{l_1,0,l_3;k} -
\frac{1}{q} \ch \cF^\infty_{l_1-1,0,l_3-1;k} \ ,
\end{equation}
for $l_1,l_3 > 0$ and $k\leq 4$.

It is interesting that the characters $\ch H_{\lambda;k}$ can still
be expressed in terms of ${\ch \mathcal F}^\infty_{\bl;k}$,
but the pattern of subtractions is at first sight obscure.
However, inverting the relations so as to express the
characters $\ch \cF_{\bl;k}^\infty$ in terms of the characters of the full
representations gives a clue as to what is happening. We find
\begin{align}
\ch {\mathcal F}^\infty_{1,1,0;4} &= \ch H_{1,1,0;4} +
\frac{1}{q} \ch H_{0,0,1;4}\nonumber \\
\ch {\mathcal F}^\infty_{1,1,1;4} &= \ch H_{1,1,1;4} +
\left(\frac{1}{q}+\frac{1}{q^2}\right)
\ch H_{0,1,0;4} + \frac{1}{q} \ch H_{2,0,0;4} +
\frac{1}{q} \ch H_{0,0,2;4}\nonumber \\
\ch {\mathcal F}^\infty_{2,1,0;4} &= \ch H_{2,1,0;4} +
\frac{1}{q} \ch H_{1,0,1;4}\nonumber \\
\ch {\mathcal F}^\infty_{1,2,0;4} &= \ch H_{1,2,0;4} +
\frac{1}{q} \ch H_{0,1,1;4} \nonumber\\
\ch {\mathcal F}^\infty_{3,1,0;4} &= \ch H_{3,1,0;4} +
\frac{1}{q} \ch H_{2,0,1;4} \nonumber\\
\ch {\mathcal F}^\infty_{1,3,0;4} &= \ch H_{1,3,0;4} +
\frac{1}{q} \ch H_{0,2,1;4} \nonumber\\
\ch {\mathcal F}^\infty_{2,2,0;4} &= \ch H_{2,2,0;4} +
\frac{1}{q} \ch H_{1,1,1;4} +
\frac{1}{q^2} \ch H_{0,0,2;4} \nonumber\\
\ch {\mathcal F}^\infty_{2,1,1;4} &= \ch H_{2,1,1;4} +
\left(\frac{1}{q}+\frac{1}{q^2}\right)
\ch H_{1,1,0;4} + \frac{1}{q} \ch H_{3,0,0;4} +
\frac{1}{q}\ch H_{1,0,2;4} + \frac{1}{q^2} \ch H_{0,0,1;4}\nonumber \\
\ch {\mathcal F}^\infty_{1,2,1;4} &= \ch H_{1,2,1;4} +
\left(\frac{1}{q}+\frac{1}{q^2}\right)
\ch H_{0,2,0;4} + \frac{1}{q} \ch H_{2,1,0;4} +
\frac{1}{q} \ch H_{0,1,2;4} + \frac{1}{q^2} \ch H_{1,0,1;4} \ .
\label{EQ:su(4)_k_decomposition}
\end{align}
In addition, we have
\begin{equation}
\ch \cF^\infty_{l_1,0,l_3;k} = \sum_{j=0}^{\min(l_1,l_3)}
\frac{1}{q^j} \ch H_{l_1-j,0,l_3-j;k} \ .
\end{equation}

All the signs on the right hand side of the above
decompositions are positive. This suggests that the
$\ch {\mathcal F}^\infty_{\bl;k}$
are indeed characters of $\widehat{su}(r+1)$ representations, but
these representations are {\it reducible\/}. We need to
understand which representations are occurring, and why the
coefficients of their characters are polynomials in
$q^{-1}$. This we will do in the next section.

\section{Kostka polynomials}
\label{sec_kospol}

In this section we will introduce Feigin and Loktev's $q$-refinement
\cite{feigin-loktev} of the Littlewood-Richardson coefficients that
occur in the decomposition of tensor products of the finite-dimensional $su(r+1)$
representations \cite{littlewood-richardson}.

Let $\omega_i$ denote the fundamental weights
of the finite-dimensional $su(r+1)$ algebra. A finite-dimensional irreducible
representation labeled by the Dynkin indices $l_i\in \ZZ_{\geq 0}$ has
highest weight $\lambda=\sum_{i=1}^{r} l_i \omega_i$ and corresponds
to a Young diagram with $l_1$ columns containing one box,
$l_2$ columns with two boxes, and so on:
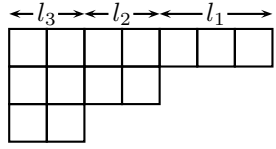
\begin{center}
\psset{unit=1mm,dimen=middle}
\begin{pspicture}(0,0)(35,20)
\multips(0,0)(5,0){2}{\psframe(0,0)(5,5)}
\multips(0,5)(5,0){4}{\psframe(0,0)(5,5)}
\multips(0,10)(5,0){7}{\psframe(0,0)(5,5)}
\rput(5,17){$l_3$}
\rput(15,17){$l_2$}
\rput(27.5,17){$l_1$}
\psset{arrows=->}
\psline(3.2,17)(0,17)
\psline(6.5,17)(10,17)
\psline(13.2,17)(10,17)
\psline(16.5,17)(20,17)
\psline(25.7,17)(20,17)
\psline(29,17)(35,17)
\psset{arrows=-}
\end{pspicture} \ .
\end{center}

The Littlewood-Richardson rules \cite{littlewood-richardson} 
provide a product that allows us to make the space of Young
diagrams into an associative algebra. The multiplication
operation in this algebra mirrors the
multiplication and decomposition of the Schur functions
corresponding to the Young diagrams, and, these being the
characters of the associated $su(r+1)$ representations,
reflect the decomposition of $su(r+1)$ tensor product
representations into their irreducible components. For
example, using the notation $V_{l_1,l_2,l_3}$ for the
$su(4)$ representation with Dynkin indices $l_1,l_2,l_3$,
the Young diagram manipulation

\medskip

\begin{center}
\psset{unit=1mm,dimen=middle,linewidth=.2}
\begin{pspicture}(0,0)(130,15)
\psframe(0,5)(5,10)
\multips(0,10)(5,0){2}{\psframe(0,0)(5,5)}
\multips(20,10)(5,0){2}{\psframe(0,0)(5,5)}
\multips(40,10)(5,0){4}{\psframe(0,0)(5,5)}
\psframe(40,5)(45,10)
\multips(70,5)(5,0){2}{\psframe(0,0)(5,5)}
\multips(70,10)(5,0){3}{\psframe(0,0)(5,5)}
\psframe(95,0)(100,5)
\psframe(95,5)(100,10)
\multips(95,10)(5,0){3}{\psframe(0,0)(5,5)}
\psframe(120,0)(125,5)
\multips(120,5)(5,0){2}{\psframe(0,0)(5,5)}
\multips(120,10)(5,0){2}{\psframe(0,0)(5,5)}
\rput(15,12.5){{\LARGE$\otimes$}}
\rput(35,12.5){{\LARGE$=$}}
\rput(65,12.5){{\LARGE$\oplus$}}
\rput(90,12.5){{\LARGE$\oplus$}}
\rput(115,12.5){{\LARGE$\oplus$}}
\end{pspicture}
\end{center}
corresponds to the decomposition
\be
V_{1,1,0}\otimes V_{2,0,0} = V_{3,1,0}\oplus V_{1,2,0}\oplus
V_{2,0,1}\oplus V_{0,1,1}.
\ee
The coefficients are not always unity. For example

\medskip

\begin{center}
\psset{unit=1mm,dimen=middle,linewidth=.2mm}
\begin{pspicture}(0,0)(120,15)
\multips(0,0)(0,5){3}{\psframe(0,0)(5,5)}
\multips(15,5)(0,5){2}{\psframe(0,0)(5,5)}
\psframe(30,10)(35,15)
\multips(45,0)(0,5){3}{\psframe(0,0)(5,5)}
\multips(50,5)(0,5){2}{\psframe(0,0)(5,5)}
\psframe(55,10)(60,15)
\multips(70,0)(0,5){3}{\psframe(0,0)(5,5)}
\multips(75,0)(0,5){3}{\psframe(0,0)(5,5)}
\multips(90,10)(5,0){2}{\psframe(0,0)(5,5)}
\multips(115,5)(0,5){2}{\psframe(0,0)(5,5)}
\rput(10,12.5){{\LARGE$\otimes$}}
\rput(25,12.5){{\LARGE$\otimes$}}
\rput(40,12.5){{\LARGE$=$}}
\rput(65,12.5){{\LARGE$\oplus$}}
\rput(85,12.5){{\LARGE$\oplus$}}
\rput(105,12.5){{\LARGE$\oplus$}}
\rput(111,12.5){{\LARGE$2$}}
\end{pspicture} \ ,
\end{center}
or
\be
V_{0,0,1}\otimes V_{0,1,0}\otimes V_{1,0,0}=
V_{1,1,1} \oplus V_{0,0,2} \oplus V_{2,0,0} \oplus 2\, V_{0,1,0} \ .
\label{EQ:littlewood2}
\ee

This last expression is to be compared with the
decomposition of our $\wsu(4)_{k=4}$ function space
\be
\ch {\mathcal F}^\infty_{1,1,1;4} =
\ch H_{1,1,1;4} +
\frac{1}{q} \ch H_{0,0,2;4} +
\frac{1}{q} \ch H_{2,0,0;4} +
\left(\frac{1}{q}+\frac{1}{q^2}\right) \ch H_{0,1,0;4} \ .
\label{EQ:kostka1}
\ee
If we set $q=1$ in the coefficients in this
decomposition, we recover the integers appearing in
(\ref{EQ:littlewood2}). All representations appearing in
(\ref{EQ:su(4)_k_decomposition}) are similarly accounted
for: The $H_{\widehat{\lambda}_i}$ appearing in the
decomposition of $\ch \cF^\infty_{l_1,l_2,l_3;4}$ are precisely the
$\wsu(4)$ representations whose top grades
are the $V_{\lambda_i}$ in the decomposition
of the product $V_{l_1,0,0}\otimes V_{0,l_2,0}\otimes V_{0,0,l_3}$,
and if $q\to 1$ their $q$-coefficients reduce to
the multiplicity of these $V_{\lambda_i}$.

We need to understand why the coefficients are $q$-dependent.
Following Feigin and Loktev \cite{feigin-loktev}, we now show that
these $q$-polynomial coefficients can be introduced even for 
finite-dimensional $su(r+1)$ representations. Let $V_{\lambda_i}$ denote a
collection of irreducible finite-dimensional highest-weight
representations of a Lie algebra $\mathfrak{g}$, and consider the
tensor product
\be
V_{\lambda_1}\otimes\cdots\otimes V_{\lambda_N}.
\ee
To each $V_{\lambda_i}$ we associate a distinct complex
number $\zeta_i$ (to be thought of as ``where'' the representation
is located). We then introduce the algebra
$\mathfrak{g}[t]\equiv \mathfrak{g}\otimes {\mathbb C}[t]$ of
$\mathfrak{g}$-valued polynomials in $t$, with Lie bracket
\be
[a \otimes t^m, b\otimes t^n]= [a,b]\otimes t^{m+n}.
\ee
(Since only non-negative powers of $t$ are being allowed, there is no
non-trivial central extension, so
this is {\it not\/} the affine Lie algebra associated with
$\mathfrak{g}$.)

The representation $V_{\lambda_i}$ gives rise to the {\em evaluation
representation} of $\mathfrak{g}[t]$ by setting 
\begin{equation}
(a \otimes t^m) \ket{u} = \zeta_i^m a \ket{u} \ , \qquad
{\rm for} \ \ket{u} \in V_{\lambda_i} \ . 
\end{equation}
We similarly define the action of 
$\mathfrak{g}[t]$ on the tensor product space.
If $a\in \mathfrak{g}$, and $\ket{u_i}\in V_{\lambda_i}$, we set
\be
\label{coprod}
(a\otimes t^m)(\ket{u_1}\otimes\cdots \otimes \ket{u_N})=
\sum_i \zeta_i^m (\ket{u_1}\otimes \cdots \otimes
a \ket{u_i} \otimes \cdots \otimes \ket{u_N}).
\ee
We will call this action the geometric, or fusion, co-product.
For $m=0$ it reduces
to the usual co-product action of $\mathfrak{g}$ on the product space
\be
a\to \Delta(a) = \sum \mathbb{I}\otimes \cdots\otimes a\otimes \cdots
\otimes \mathbb{I} \ .
\ee

If $\ket{\lambda_i}$ are the highest weight
(and hence cyclic) vectors for the representations
$V_{\lambda_i}$, then, with the usual co-product
action of $\mathfrak{g}$, the vector
$\ket{v}\equiv \ket{\lambda_1}\otimes \cdots \otimes \ket{\lambda_N}$
is the highest weight vector for only one of
the irreducible representations occurring in the decomposition of the
tensor product of the $V_{\lambda_i}$. Under the action
of $\mathfrak{g}[t]$, however, and provided that the $\zeta_i$
are all distinct, $\ket{v}$ is a cyclic vector for the entire tensor
product space. This is because the matrix $\zeta_i^j$ is invertible,
and so any vector
$\ket{u_1}\otimes\cdots \otimes a\ket{u_j}\otimes \cdots\otimes \ket{u_N}$
is obtainable as a linear combination of
$(a\otimes t^m)(\ket{u_1}\otimes\cdots \otimes \ket{u_N})$
for different $m$, and any vector $\ket{u_i}$ in each of the
$V_{\lambda_i}$ is obtainable by the action of suitable $a$ on
$\ket{\lambda_i}$.

The Lie algebra $\mathfrak{g}[t]$ is graded by the degree of
the polynomial in $t$. This grading extends to the universal enveloping
algebra $U(\mathfrak{g}[t])$ and gives rise to a
{\it filtration\/} --- a nested set of
vector spaces ---
\be
F^0\subseteq \ldots \subseteq F^{i}\subseteq
F^{i+1}\subseteq \ldots,
\ee
where
\be
F^i= U^{\leq i}(\mathfrak{g}[t])\ket{v}.
\ee
Here, $U^{\leq i}(\mathfrak{g}[t])$ denotes the elements of the
universal enveloping algebra $U(\mathfrak{g}[t])$, which have
degree in $t$ less than or equal to $i$.
The action of $\mathfrak{g}$, considered as the zero-grade
component of $\mathfrak{g}[t]$, preserves this
filtration and so has a well defined action on each of the
components of the 
associated graded spaces ${\rm gr}^i[F]\equiv F^i/F^{i-1}$. Each of the
irreducible representations in the tensor product space will
appear in one of these graded subspaces.

As an illustration, consider the simplest case $\mathfrak{g}=su(2)$
where
\be
[e,f]=h, \quad [h,e]=+2e,\quad [h,f]=-2f.
\ee
In the usual physics spin-$j$ notation, we have the
decomposition $\tsh\otimes \tsh = 1\oplus 0$, or in our
Dynkin index language
\be
V_1\otimes V_1 =V_2\oplus V_0.
\ee
The repeated action of the step down operator $f$
on the state
\be
\ket{v}=\ket{\ua}\otimes\ket{\ua}
\ee
(where $\ket{\ua}$ and $\ket{\da}$ denote the states of weight $1$
and $-1$ respectively)
generates only the three states appearing in the spin-$1$
representation $V_2$, which therefore constitutes the space
$F^0$. The action of $f \otimes t$, however, yields
\bea
(f \otimes t)\ket{v} &=& \zeta_1 \ket{\da}\otimes \ket{\ua} +
\zeta_2 \ket{\ua}\otimes \ket{\da},\nonumber\\
&=& \tsh({\zeta_1-\zeta_2})\left(\ket{\da}\otimes \ket{\ua} -
\ket{\ua}\otimes \ket{\da}\right) \nonumber\\
&&\quad + \tsh({\zeta_1+\zeta_2})\left(\ket{\da}\otimes
\ket{\ua} + \ket{\ua}\otimes \ket{\da}\right),
\eea
which, since
$\ket{\da}\otimes \ket{\ua} + \ket{\ua}\otimes \ket{\da}$
lies in $F_0$, is equivalent in $F^1/F^0$ to
\be
 \tsh({\zeta_1-\zeta_2})\left(\ket{\da}\otimes \ket{\ua} - \ket{\ua}\otimes
\ket{\da}\right).
\ee
This last vector is the highest weight (indeed the only
weight) in the spin-$0$ representation.

While the highest-weight vectors of each representation
occurring in the tensor product must appear somewhere in
this construction, their coefficients depend quite
non-trivially on the $\zeta_i$, and some of these coefficients
might vanish at non-generic points 
even for non-coincident $\zeta_i$. It is
therefore not obvious that the grade at which a given
representation first appears is independent of the choice
of the $\zeta_i$. Feigin and Loktev conjecture that this is
the case, and if this is true, the graded character
\be
\ch_q (V_{\lambda_1}\otimes \cdots \otimes V_{\lambda_N})
= \sum_d q^d \ch ({\rm gr}^d[F]) 
\ee
should be independent of the $\zeta_i$.
In \cite{usII}, we provide a proof of this conjecture by Feigin and
Loktev in the spacial case where $\mathfrak{g}= su(r+1)$ and when the 
$\lambda_i$ correspond to rectangular representations.

In the $su(2)$ example the $\zeta_i$ independence is manifest,
and we have
\be
\ch_q(V_1\otimes V_1) =\ch V_2 + q \ch V_0.
\ee
Applying this construction to $su(4)$ we would find
\be
\ch_q (V_{0,0,1}\otimes V_{0,1,0}\otimes V_{1,0,0}) =
\ch V_{1,1,1} + (q+ q^2)
\ch V_{0,1,0} + {q} \ch V_{2,0,0} + {q}
\ch V_{0,0,2},
\ee
which looks like (\ref{EQ:kostka1}), but with
$q\leftrightarrow 1/q$ to reflect the fact that Feigin and Loktev
define their grading in the opposite direction to the one which is
natural for the affine algebra.

To obtain the characters for arbitrary representations of $\wsu(r+1)$,
we only need to consider the fusion product of $r$ rectangular
representations $V_1,\ldots,V_r$, such that the highest weight
$\lambda_i$ of $V_i$ is given by $\lambda_i = n_i \omega_i$, with
$n_i \geq 0$. We will denote the $q$-polynomial coefficients in the
decomposition of the character of this fusion product by
$\cK_{\bl,\bn} (q)$. Here, $\bl$ is the vector whose entries are the
Dynkin indices of the representations present in the fusion
product. The entries of $\bn$ are the $n_i$. In appendix
\ref{app_kostka}, we show how these $q$-polynomials can be calculated.
The result is the fermionic formula \eqref{kospol}. Details can
be found in our paper \cite{usII}, where we also showed that the
$q$-polynomials are in fact generalized Kostka polynomials
\cite{SchWar,KirShi}. Note that we will always
assume that $\sum_i n_i \leq k$, so we do not impose the level-restriction
conditions in calculating the Kostka polynomials%
\footnote{Level-restricted generalized Kostka polynomials \cite{SchShi}
are also important, but do not play a role in our calculations.}.
This restriction is always met in our case of obtaining characters of
general $\wsu(r+1)$ representations.

Thus, we have the following expression for the character of the fusion product
of rectangular representations
\begin{equation}
\label{fusproddecomp}
\ch_q (V_1 \otimes V_2 \otimes \cdots \otimes V_r) =
\sum_{\bl} \cK_{\bl,\bn} (q) \ch V_\bl \ .
\end{equation}

These $q$-polynomial $\cK_{\bl,\bn}(q)$ coefficients that reduce for
$q=1$ to the Littlewood-Richardson coefficients are generalized Kostka
polynomials. As a result, we conclude that the sum in equation
\eqref{fusproddecomp} is finite.

Classical Kostka polynomials are parametrized by two Young diagrams,
$\lambda$ and
$\mu$, such that $|\lambda| = |\mu|$. Generalized Kostka polynomials
\cite{SchWar}, which we consider in this paper, are parametrized by a
Young diagram $\lambda$ and a set of rectangular diagrams $\{\mu_i\}$,
such that the total number of boxes in the set $\{\mu_i\}$ is the same
as the number of boxes in $\lambda$. Note that these Young diagrams are
associated to $\mathfrak{gl}_{r+1}$. In our notation used above, the
Young diagrams are associated to $\mathfrak{sl}_{r+1}$, which can be
obtained from the $\mathfrak{gl}_{r+1}$ diagrams by `stripping off'
the columns of height $r+1$ (see \cite{usII} for more details).

The classical Kostka polynomial is a special case of the
generalized Kostka polynomial, where the set $\{\mu_i\}$ 
case is the set of single-row diagrams, each equal to a single
row of the diagram $\mu$.

In both cases, the integer $K_{\lambda,\{\mu_i\}}(1)$ is equal to the
multiplicity of the $\gl_n$-module $V_\lambda$ in the tensor product
of representations $V_{\mu_1}\otimes\cdots \otimes V_{\mu_N}$. The Kostka
polynomial is therefore a refinement of tensor product
multiplicities, or a grading.

\section{Wess-Zumino-Witten conformal theory}
\label{sec_wzwcft}

Although the geometric, or fusion, co-product and the evaluation
representation were initially defined for the finite-dimensional Lie algebra
$\mathfrak{g}$, they are motivated by the action of the
affine Lie algebra $\widehat {\mathfrak{g}}$ in
Wess-Zumino-Witten (WZW) conformal field theory \cite{WZW}.

We will focus on the holomorphic half of a level-$k$ WZW model defined
on the Riemann sphere. The
symmetry algebra of this model is $\mathfrak{g}\otimes {\mathcal
  M}(\boldsymbol\zeta)$, consisting of $\mathfrak{g}$-valued
meromorphic functions with possible poles at the points $\zeta_i$
(where $\boldsymbol\zeta = (\zeta_1,...,\zeta_N)$).

Let $\lambda$ denote the highest weight of
a finite-dimensional representation $V_\lambda$ of the finite-dimensional
algebra $\mathfrak{g}$,
and suppose that the corresponding (in that its top graded component
is $V_\lambda$) representation
$H_{\widehat{\lambda}}$ of the level-$k$ affine Lie
algebra $\widehat{\mathfrak{g}}$ is integrable.
Then, the Wess-Zumino primary field
$\varphi_{\lambda}(\zeta)$ acts on the vacuum at
the origin $\zeta=0$ there to create the highest weight state
$\ket{\widehat{\lambda}}= \varphi_\lambda (0) \ket{0}$ of
$H_{\widehat{\lambda}}$.
If
\be
a(z)=\sum_{n=-\infty}^{\infty} a[n] z^{-n-1}
\ee
is the WZW current associated with the element $a\in
\mathfrak{g}$, and $\gamma$ is any contour surrounding the
origin then
\be
a[n] = \frac 1{2\pi i}\oint_\gamma z^n a(z) \,dz
\ee
has its usual action as an element of
$\widehat{\mathfrak{g}}$ on this representation.

More generally, for integrable
$\widehat{\lambda}_i$ we create highest
weight states
$\ket{{\widehat{\lambda}_i}}_{\zeta_i}$ at the
points $\zeta_i$ by acting on the vacuum at these points with
$\varphi_{\lambda_i}(\zeta_i)$ on the vacuum at $\zeta_i$. The current
algebra acts on the tensor product
$H_{\widehat{\lambda}_1}\otimes \cdots
\otimes H_{\widehat{\lambda}_N}$ of these
spaces. We recall how this comes about. Let $f(t)$ denote a
function meromorphic on the Riemann sphere, and having
poles at no more than
the $\zeta_i$. If
$\Gamma$ is a contour surrounding $\zeta_1,\ldots, \zeta_N$, then
we insert $\frac 1{2\pi i} \oint_\Gamma f(t) a(t)\, dt$
into a suitable correlation function. The contour can be
deformed to a sum of contours $\gamma_i$ each enclosing
only one of the $\zeta_i$, and the $a [n]$ acting on the
representation space at $\zeta_i$ are the local mode-expansion
coefficients $a(z)=\sum_n a[n](z-\zeta_i)^{-n-1}$, or
\be
a [n]= \frac{1}{2\pi i} \oint (t-\zeta_i)^n a(t)\, dt.
\ee
Consequently,
if
\be
f(t) = \sum_{n = -p_i}^\infty f_n(\zeta_i) \,(t-\zeta_i)^n
\ee
is the Laurent expansion of $f$ about $\zeta_i$ then
$a\otimes f$ acts on a state
$\ket{v}_{\zeta_i}\in H_{\lambda_i}$ as
\be
\sum_{n=-p_i}^\infty f_n(\zeta_i) \,a[n]
\ket{v}_{\zeta_i},
\ee
and on $H_{\widehat{\lambda}_1}\otimes \cdots \otimes
H_{\widehat{\lambda}_N}$ as
\be
\Delta_{\zeta_1,\ldots,\zeta_N}(a\otimes f)(\ket{v_1}_{\zeta_1}\otimes
\cdots\otimes \ket{v_N}_{\zeta_N}) =
\sum_i \ket{v_1}_{\zeta_1}\otimes \cdots\otimes
\left(\sum_{n=-p_i}^\infty f_n(\zeta_i) \,a[n]
\ket{v_i}_{\zeta_i}\right)\otimes \cdots \otimes\ket{v_N}_{\zeta_N},
\ee
where $\Delta_{\zeta_1,\ldots,\zeta_N}$ is the
``geometric'' co-product \cite{moore-seiberg_co-product}
\be
\Delta_{\zeta_1,\ldots,\zeta_N}(a\otimes f) =
\sum_i \mathbb{I}\otimes\cdots \otimes
\left(\sum_{n=-p_i}f_n(\zeta_i)
\,a[n]\right)\otimes \cdots \otimes \mathbb{I}.
\ee
This co-product makes the tensor products of any
number of level-$k$ representations into a level-$k$
representation. With the conventional co-product, the level
would be $Nk$.

In the particular case
that $f(t)=t^n$, the Laurent expansion about $\zeta_i$ is
\be
t^n=\sum_{m=0}^n \binom{n}{m} \zeta_i^{n-m}(t-\zeta_i)^m.
\ee
If $\ket{v}_{\zeta_i}$ is a top-component state on which $a[n]$ with
$n>0$ acts as zero, the action on $\ket{v}_{\zeta_i} $ is as
$\zeta_i^n a[0]\ket{v}_{\zeta_i}$.
Since the zero-grade elements of $\widehat{\mathfrak{g}}$
form an algebra isomorphic to $\mathfrak{g}$, we recognize
our evaluation co-product action \cite{feigin-loktev} described in
equation \eqref{coprod}.

When the contour $\Gamma$ encloses all insertions of
$\varphi_{\lambda}$'s
on the Riemann sphere, it can be contracted away, and the algebra acts
trivially.
This observation motivates the definition of the space
${\mathcal H}(\bzeta, \{\lambda\})$ of {\it conformal
blocks\/}. A conformal block $\Psi\in {\mathcal
H}(\bzeta;\{\lambda\})$ is a mapping
\be
\Psi: H_{\widehat \lambda_1}\otimes \cdots \otimes H_{\widehat\lambda_N}
\to {\mathbb C} \ ,
\ee
invariant under the action of any $a\otimes f$, $a\in
\mathfrak{g}$, with $f$ having poles at no more that the
$\zeta_i$. It may be shown \cite{kohno} that such a mapping
is uniquely specified by its evaluation on the top graded component
$V_{\lambda_1}\otimes \cdots \otimes V_{\lambda_N}$, and that the
dimension
${\rm dim\,}\bigl({\mathcal H}(\bzeta,\{\lambda\})\bigr)$ of
the space of conformal blocks is the multiplicity of the
identity operator in the fusion product of the
$\varphi_{\lambda_i}$.

This definition coincides with the physicist's picture where
\be
\Psi(\zeta_1,\ldots,\zeta_N)\equiv
\Psi\left(\ket{\mu_1}_{\zeta_1} \otimes \cdots
\otimes \ket{\mu_N}_{\zeta_N}\right)=
\eval{0}{R\{\varphi_{\mu_1}(\zeta_1)\ldots
\varphi_{\mu_N}(\zeta_N)\}}{0},
\ee
with $\mu_i$ weights in $V_{\lambda_i}$, is  thought of as a
radial-ordered correlation function of the primary fields
$\varphi_{\mu_i}$.  This correlator is not uniquely defined because
the chiral primary fields are not mutually local. It is, however, a
solution of the Knizhnik-Zamolodchikov equations, whose solution space
is precisely the space of conformal blocks as identified above.  In
the operator language this comes about because the
$\bra{0}$ could be the dual of any of the copies
of the highest-weight state of the vacuum representation that occur in 
the
fusion-product.

We now recognize what it is that our function-space character is
counting. It is the number of linear independent functions of the form
\bea
\label{funcspace}
F(\sz)&=&\eval{v}
{R\{f_{\alpha_1} (z_{1}^{(1)})\ldots f_{\alpha_r}(z_{m^{(r)}}^{(r)})
\varphi_{\lambda_1}(\zeta_1)\ldots
\varphi_{\lambda_r}(\zeta_r)\}}{{0}},\nonumber\\
&\equiv& \sum \Psi \left(\ket{v^*}_\infty \otimes  \ket{v_1}_{\zeta_1} 
\otimes \cdots
\otimes \ket{v_N}_{\zeta_N}\right).
\eea
Here the  $\varphi_{\lambda_i}(\zeta_i)$ create rectangular
representations with highest weight $\lambda_i=l_i \omega_i$ (no sum on 
$i$) by acting on the vacuum at the points
$\zeta_i$. The state $\bra{v}$,  analogous to the $\bra{v}$ of 
(\ref{matelem}),   is  the dual of $\ket{v}$, which lies in a   weight 
space in one of the   integrable representations 
$H_{\widehat{\lambda}}$ occurring in the fusion product. The sum is 
over the  $\ket{v_j}$  produced by the action of the $f_{\alpha}$ on 
the  $\ket{\widehat{\lambda}_i}$.  The state     $\ket{v^*}_\infty$  is 
in the dual representation  $H_{\widehat{\lambda}}^*$ that is located 
at the point $\infty$ on the Riemann sphere. The  (left action)  
$\mathfrak{g}$ weight of $\ket{v^*}$ is   minus that  of $\ket{v}$.  
(The correspondence $\bra{v}\leftrightarrow \ket{v^*}_\infty$ is 
explained in \cite{kohno}.)  

The fusion product of primary fields  in the WZW model coincides with 
the finite-dimensional algebra Littlewood-Richardson rules only when $k$ is 
infinite. For finite $k$ we must use the level-restricted fusion rules 
of the Verlinde algebra.  The level restrictions have no effect, 
however, when we build up an integrable representation of   $\widehat 
{\mathfrak{g}}$ by concatenating integrable rectangular
representations, as we are doing in this paper. The representations 
appearing in our  WZW fusion product therefore coincide with those 
appearing in the $\mathfrak{g}[t]$ fusion product, which in 
turn are given by the Littlewood-Richardson rules of $\mathfrak{g}$.

The fusion product itself is not a graded space. However, we can
define a filtration on this space, in the same way as in the
Feigin-Loktev fusion product. This filtration is inherited from the
filtration of the algebra $U(\n_-\otimes\C(t^{-1}))$, where we assign a
degree zero to the product of primary fields. The associated graded
space has the character \eqref{finfchar}, which we can regard as the
character of the fusion product.  It is a sum over irreducible
characters with a certain shift in the overall degree. We can regard
this as the statement that the vector $\bra v$ in the matrix element
belongs to a highest weight representation, with a highest weight on
which $d$ acts by some negative integer instead of 0. (For an
alternative point of view, where the space of highest weight vectors
is interpreted as a quotient of the integrable representation -- which
is naturally graded -- see \cite{FJKLM}.)

We would like to stress again that the
characters of this space do not depend on the locations $\zeta_i$ of
the rectangular representations!

\section{Characters for arbitrary $\wsu(r+1)_k$ representations}

In this section, we will combine the results of the previous section
and give explicit character formul{\ae} for arbitrary (integrable)
representations of $\wsu(r+1)_k$. There is one more ingredient needed
to to this, namely an explicit formula for the (generalized) Kostka
polynomials, which form an essential ingredient in the character
formul{\ae}, as explained before. In this section, we will
give the character formul{\ae} in terms of the generalized Kostka
polynomials. How to obtain an explicit formula will be
described in appendix \ref{app_kostka} (we refer to \cite{usII} for
the details). The resulting formula for the generalized Kostka
polynomials is stated in equation \eqref{kospol}.

We can write the decomposition of the
character $\cF_{\bn;k}^{\infty} (q,\bx)$ in the following way.
\begin{equation}
\label{fpdecomp}
\ch \cF_{\bn;k}^\infty (q,\bx) = \sum_{\bl} \cK_{\bl,\bn}
\left(\frac{1}{q}\right) \ch H_{\bl;k} (q,\bx) \ .
\end{equation}
We need to make a few remarks about the sum in the decomposition
\eqref{fpdecomp}. First of all, we would like to note that this sum is
finite. To show this, we introduce the notion of the threshold level,
which is the lowest level for which an highest weight $\lambda$
corresponds to a highest weight representation. We will denote this
threshold level by $k(\bl)$. For $\wsu(r+1)$, it is simply given by
$k(\bl) = \sum_{i=1}^{r} l_i$. The only $\bl$ for which the Kostka
polynomial $\cK_{\bl,\bn}(q)$ is non-zero, are the $\bl$ such that
$k(\bl) \leq k(\bn)$. The only non-zero $\cK_{\bl,\bn} (q)$ with
$k(\bl) = k(\bn)$ is when $\bl=\bn$, in which case
$\cK_{\bn,\bn}(q)=1$, see equation \eqref{kosprop}.
Using these results,
we can view the Kostka polynomials $\cK_{\bn,\bl} (q)$ as elements of
a square matrix $\bK$, which is upper triangular and $1$'s on the diagonal.
Hence, this matrix is invertible, if we
order the `highest weights' $\bl$ according to increasing
threshold level. Note that we did not specify an ordering for weights
with the same threshold level, but any ordering of those weights gives
an upper triangular matrix.

Now that we established that the Kostka matrix $\bK$ is
invertible, we can invert the relation \eqref{fpdecomp} to obtain
explicit character formul{\ae} for arbitrary (integrable) highest
weight representations of $\wsu(r+1)_k$
\begin{equation}
\label{chahw}
\ch H_{\bl;k} (q,\bx) = \sum_{\bn} (\bK^{-1})_{\bn,\bl}
\left(\frac{1}{q}\right)
\ch \cF_{\bn;k}^\infty (q,\bx)  \ .
\end{equation}
Again, the (finite) sum is over highest weights $\bn$ with threshold level
$k(\bn)\leq k(\bl)$.

\section{Conclusion}

We have shown how the introduction of the fusion product
representation allows us to generalize the Feigin-Stoyanovsky strategy
for computing characters of affine $\wsu(r+1)$ to all integrable
representations. We found that for non-rectangular representations,
the character is not of fermionic type, but can be written as a linear
combination of fermionic characters. The coefficients are polynomials
in $1/q$ and are related to the generalized Kostka polynomials.

It is rather straightforward to generalize the Feigin-Stoyanovsky
construction to arbitrary affine Lie algebras. In that case, even
characters of rectangular representations are not always of
fermionic type. It turns out that the character of those
representations can be written in terms of characters of the so called
`Kirillov-Reshetikhin' modules \cite{KR90}. Details will be provided in a
forthcoming publication.

\section{Acknowledgments}

The work of EA was sponsored in part by NSF
grants DMR-04-42537 and DMR-01-32990, the work of MS by NSF grant
DMR-01-32990 and the work of RK by NSF grant DMS-05-00759.

\appendix

\section{The structure of the character of the principal subspace}
\label{app_pss}

In this appendix, we will explain the structure of the character of
the principal subspace of rectangular representations of $\wsu(r+1)$,
eq. \eqref{psschar}. To this end, we will explain the zeros of the
function $f(z)$ which appears in \eqref{ratfunc}.

The exponent of $q$ in the character is the total degree of the
polynomials $f(\sz)$, which ensure that the vanishing conditions hold,
combined with the explicit poles in $F(\sz)$. The strategy presented
in \cite{usI} and \cite{usII}, which we will outline here briefly, is
to find the minimal amount of zeros necessary to enforce the
vanishing conditions.

Let us first focus on the case we have $2 k$ variables. We need
to find a polynomial which does not vanish when $k$ variables are set
to the same location, but does vanish when $k+1$ variables are set to
the same location. Of course, the condition \eqref{intc} allows for
functions which vanish when less than $k+1$ variables are set to the
same location, but we will deal with those later.
A symmetric polynomial in the variables $z_1,z_2,\ldots,z_{2k}$
with the lowest possible degree which satisfies this property is
($\cS$ denotes symmetrization over all variables)
\begin{equation}
\label{f2k}
\cS \Bigl[
(z_1-z_{k+1})(z_2-z_{k+1})(z_2-z_{k+2})\cdots
(z_k-z_{2k-1})(z_k-z_{2k})(z_1-z_{2k}) \Bigr] \ .
\end{equation}
We can display these zeros nicely in a graphical way in terms of a
Young diagram. The boxes on the top row correspond to the variables
$z_1,\ldots,z_k$ while the boxes on the second row correspond to
$z_{k+1},\ldots,z_{2k}$. A line connecting the boxes of $z_i$ and
$z_j$ corresponds to the factor $(z_i-z_j)$:
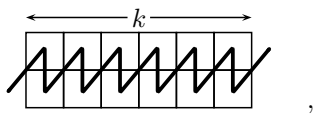
\begin{center}
\psset{unit=1mm}
\begin{pspicture}(-5,0)(35,15)
\psset{dimen=middle,linewidth=.2mm}
\multips(0,0)(5,0){6}{\psframe(0,0)(5,5)}
\multips(0,5)(5,0){6}{\psframe(0,0)(5,5)}
\rput(15,12){$k$}
\psline[arrows=->](13.5,12)(0,12)
\psline[arrows=->](16.5,12)(30,12)
\psset{linewidth=.5mm,arrows=cc-cc}
\psline[linearc=.1](-2.5,2)(2.5,8)(2.5,2)(7.5,8)(7.5,2)(12.5,8)
(12.5,2)(17.5,8)(17.5,2)(22.5,8)(22.5,2)(27.5,8)(27.5,2)(32.5,8)
\psset{arrows=-}
\end{pspicture} \ ,
\end{center}
where `periodic boundary conditions' are assumed, to obtain the zero
$(z_1-z_{2k})$. It easily follows that if we pick $k$ variables, and
set all of them to the same value, there will always be a term in the
symmetrization of eq. \eqref{f2k} which is not zero. However, setting
$k+1$ variables to the same location gives a zero for each term in the
symmetrization.

We will now discuss the general case, {\it i.e.\/} we allow for functions
which vanish when fewer than $k+1$ variables are set equal, and we
also allow for a general number of variables.

To do this, we need to label the polynomials by $r$ partitions (or
Young diagrams), one for each type of variable. These are partitions
of the integers $m^{(i)}$, with the property that the width of the
rows is maximally $k$, because the polynomials should vanish when
$k+1$ variables are set to the same value.

The number of rows of width $a$ of partition $(i)$ is given by
$\mi_a$, {\it i.e.\/} the partitions have the form
$(k^{\mi_k},(k-1)^{\mi_{k-1}},\ldots,2^{\mi_2},1^{\mi_1})$. It follows
that we have the relation $\mi = \sum_{a=1}^k a \mi_a$.  See figure
\ref{fig_partition} for an example where $m^{(1)} = 13$.

To each box of the partition $(i)$, we associate a variable $\zi_j$.
Let $z_1,\ldots,z_a$ be the variables associated to a row of length
$a$, of partition $(i)$ and $\bar{z}_1,\ldots,\bar{z}_{a'}$ to a row of
length $a'$, also from partition $(i)$, such that $a'\leq a$.

We need to assign the following zeros corresponding to these
variables
$$
(z_1-\bar{z}_1)(z_2-\bar{z}_1)(z_2-\bar{z}_2)(z_3-\bar{z}_2)
\cdots(z_{a'}-\bar{z}_{a'})(z_{a'+1}-\bar{z}_{a'}) \ .
$$
We will identify $z_{a+1} = z_1$. Pictorially, we can represent
these zeros as
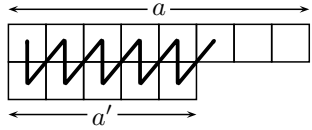
\begin{center}
\psset{unit=1mm}
\begin{pspicture}(0,-4)(40,14)
\psset{dimen=middle,linewidth=.2mm}
\multips(0,0)(5,0){5}{\psframe(0,0)(5,5)}
\multips(0,5)(5,0){8}{\psframe(0,0)(5,5)}
\psset{linewidth=.5mm,arrows=cc-cc}
\psline[linearc=.1](2.5,8)(2.5,2)(7.5,8)(7.5,2)(12.5,8)(12.5,2)(17.5,8)
(17.5,2)(22.5,8)(22.5,2)(27.5,8)
\psset{arrows=->,linewidth=.2mm}
\rput(20,12){$a$}
\psline(18.5,12)(0,12)
\psline(21.5,12)(40,12)
\rput(12.5,-2){$a'$}
\psline(10.5,-2)(0,-2)
\psline(14.5,-2)(25,-2)
\psset{arrows=->}
\end{pspicture} \ .
\end{center}
Again, each box corresponds to a variable, and a line connecting two
boxes indicates a zero when the two corresponding variables have the
same value. We obtain all the zeros needed to satisfy the
integrability condition by multiplying all the zeros associated to
each pair of rows belonging to the same partition
\begin{equation}
\prod_{i=1}^{r}\prod_{\substack{{\rm pairs\ of\ rows}\\
{\rm of\ partition}\ (i) \\ {\rm of\ length}\ a,a'\\a\geq a'}}
\prod_{j=1}^{a'} (\zi_j-\bar{z}^{(i)}_j)(\zi_{j+1}-\bar{z}^{(i)}_j) \ .
\end{equation}
To clarify this, we will look at one particular variable (which will
correspond to the black box in the diagram below), and show with which
other variables it has a zero. These other variables are given by
the gray boxes:
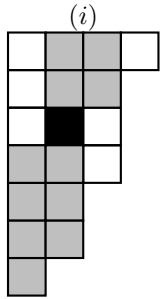
\begin{equation}
\label{intcp}
\psset{unit=1mm,dimen=middle}
\begin{pspicture}(0,0)(20,40)
\multips(0,20)(0,5){3}{\psframe(0,0)(5,5)}
\psframe(15,30)(20,35)
\psframe(10,15)(15,20)
\psframe(10,20)(15,25)
\psset{fillstyle=solid,fillcolor=black}
\psframe(5,20)(10,25)
\psset{fillcolor=lightgray}
\multips(0,0)(0,5){4}{\psframe(0,0)(5,5)}
\multips(5,5)(0,5){3}{\psframe(0,0)(5,5)}
\multips(5,25)(0,5){2}{\psframe(0,0)(5,5)}
\multips(10,25)(0,5){2}{\psframe(0,0)(5,5)}
\psset{fillstyle=none}
\rput(10,37){$(i)$}
\end{pspicture} \ .
\end{equation}
To obtain all the zeros, we need to take the product over all `black
boxes', but without double counting the zeros. Finally, we need to
take the product over all partitions $i=1,\ldots,r$.
It is not to hard to convince oneself that these zeros do imply the
integrability condition.

We will now focus on the zeros we have to include in the polynomials
$f(\{(\zi_j\})$ to ensure that the Serre relations are taken into account
properly. Let $\zi_j$ be a variable associated to a row of length $a$
of the partition of $m^{(i)}$. This variable has zeros with variables
$\zii_{j'}$ which belong to a row of the partition of $m^{(i+1)}$ of
length $a'$. In particular, these zeros are
$\prod_{j'\neq j} (\zi_j-\zii_{j'})$.
We obtain all the zeros associated to the pair of partitions $(i)$ and
$(i+1)$ by multiplying all the zeros associated to each variable of
partition $i$.
\begin{equation}
\prod_{i=1}^{r-1}\prod_{\substack{{\rm rows\ of}\\{\rm partition}\ (i)}}
\prod_{\substack{{\rm rows\ of}\\{\rm partition}\ (i+1)}}
\prod_{\substack{j,j'\\j'\neq j}}
(\zi_j-\zii_{j'}) \ .
\end{equation}
Pictorially, these zeros are given by
\begin{equation}
\label{serrecp}
\psset{unit=1mm,dimen=middle}
\begin{pspicture}(0,0)(65,40)
\multips(0,0)(0,5){7}{\psframe(0,0)(5,5)}
\multips(5,5)(0,5){4}{\psframe(0,0)(5,5)}
\psframe(5,30)(10,35)
\multips(10,20)(0,5){3}{\psframe(0,0)(5,5)}
\psframe(15,30)(20,35)
\psset{fillstyle=solid,fillcolor=black}
\psframe(5,25)(10,30)
\psset{fillcolor=lightgray}
\multips(30,5)(0,5){6}{\psframe(0,0)(5,5)}
\multips(40,15)(0,5){4}{\psframe(0,0)(5,5)}
\multips(45,20)(0,5){3}{\psframe(0,0)(5,5)}
\multips(50,25)(0,5){2}{\psframe(0,0)(5,5)}
\psset{fillstyle=none}
\multips(35,10)(0,5){5}{\psframe(0,0)(5,5)}
\rput(10,37){$(i)$}
\rput(42.5,37){$(i+1)$}
\end{pspicture} \ .
\end{equation}
Again, we need to take the product over all `black boxes' and over
$i=1,\ldots,r-1$. Again, it is easy to check that these zeros,
combined with the zeros which enforce integrability, give rise to the
Serre conditions.

Finally, we need to include the zeros to make sure the highest weight
condition is satisfied. For a row of length $a>l$ of the partition of
$m^{(p)}$, we need to include the zeros
\begin{equation}
\prod_{{\substack{{\rm rows\ of}\\{\rm partition}\ p}}}
\prod_{j> l}\zp_j \ ,
\end{equation}
or, pictorially, the zeros are given by the gray boxes
\begin{equation}
\label{topcp}
\psset{unit=1mm,dimen=middle}
\begin{pspicture}(0,5)(30,35)
\multips(0,5)(0,5){5}{\psframe(0,0)(5,5)}
\multips(5,10)(0,5){4}{\psframe(0,0)(5,5)}
\multips(10,10)(0,5){4}{\psframe(0,0)(5,5)}
\psset{fillstyle=solid,fillcolor=lightgray}
\multips(15,20)(0,5){2}{\psframe(0,0)(5,5)}
\multips(20,20)(0,5){2}{\psframe(0,0)(5,5)}
\psframe(25,25)(30,30)
\psset{fillstyle=none,arrows=->}
\rput(7.5,32){$l$}
\rput(25,32){$(p)$}
\psline(6,32)(0,32)
\psline(9,32)(15,32)
\end{pspicture} \ .
\end{equation}

Counting the total degree of the zeros implied by
\eqref{intcp}, \eqref{serrecp} and \eqref{topcp}, combined with the
degree corresponding to the poles in \eqref{ratfunc} and the extra
term $\sum_{i=1}^r \mi$ in the definition \eqref{funchar} precisely
give the exponent of $q$ in the character formula \eqref{psschar}.

Apart from the zeros discussed above, $f(\{\zi_j\})$ can contain
additional symmetric polynomials.
The degree of these symmetric polynomials is
taken into account by the factors $\frac{1}{(q)_m}$ in
\eqref{psschar}, see \cite{usII} for the details. The factor
$\frac{1}{(q)_m}$ is the generating function for the symmetric
polynomials in $m$ variables. The coefficient of $q^d$ in the
expansion of $\frac{1}{(q)_m}$ is the number of symmetric polynomials
of degree $d$ in $m$ variables.

\section{Obtaining the Kostka polynomials}
\label{app_kostka}

In the main text, we explained how the characters of the fusion
product of rectangular representations can be decomposed into
characters of arbitrary highest-weight representations. In this
appendix we will outline a strategy to obtain an explicit expression
for the `expansion coefficients', which turn out to be generalized
Kostka polynomials. We refer to our paper \cite{usII} for the details
of the proof.

We need to consider matrix elements of the form
\begin{equation}
\label{matel}
G_{\lambda,\bmu} (\bzeta) =
\eval{u_{\lambda}}
{f_{\alpha_1}(z^{(1)}_1) \cdots f_{\alpha_r}(z^{(r)}_{m^{(r)}})}
{v_{1}(\zeta_1)\otimes\cdots\otimes v_{r}(\zeta_r)} \ ,
\end{equation}
where $\bra{u_{\lambda}}$ is dual to the state
$\ket{u^*_\lambda}_\infty$, as explained in the discussion following
equation \eqref{funcspace}.
The $v_i (\zeta_i)$ are the highest weights of the rectangular
representations $H_{n_i \omega_i} (\zeta_i)$.  Thus, the
rectangular representations inserted at the points $\zeta_i$ are given
by $n_i \omega_i$. We define $\bn=(n_1,\ldots,n_r)^T$.  In
addition, we define $\bmu=\sum_i n_i \omega_i$ and the coefficients $l_i$
are determined by $\lambda = \sum_i l_i \omega_i$.

To obtain the a formula
for the generalized Kostka polynomials, we should only consider the
bra's $\bra{u_{\lambda}}$ at infinity, which are the highest weights
of the `bottom component' of the representation located at infinity.
Note that the representation which is located at infinity is turned
`upside down', because we chose $\tfrac{1}{z}$ as the local variable
at infinity.

We need to find the conditions such that the matrix elements
\eqref{matel} are non-zero. First of all, we need that
$\lambda = \bmu-\sum_i \mi \alpha_i$, which translates into
\begin{equation}
\bm = C_r^{-1} \cdot (\bn-\bl) \ .
\end{equation}
As usual, we also need to impose the highest-weight conditions
$f_{\alpha_i}^{n_i+1} [0] \ket{v_i} = 0$ and the Serre relations.
Finally, we need to incorporate a new ingredient, which is a degree
restriction on the functions in the function space. These degree
restrictions can be derived by letting the $f$'s in the matrix elements
\eqref{matel} act to the left on $u_{\lambda}$. Now, $f_\alpha [n]$
acts trivially at $\infty$ if $n\leq 0$, which translates a degree
restriction on the functions.

The space of functions which we need to consider is spanned by
rational functions in the variables $\zi_j$, which are symmetric
under the exchange $\zi_j \leftrightarrow \zi_{j'}$. There can be
poles of order one at the positions $\zi_j = \zii_{j'}$ and at
$\zi_j = \zeta_i$. That is
\begin{equation}
\label{kosratfunc}
G(\bz) = \frac{g(\bz)}{\prod_{i=1}^{r}\prod_j (\zeta_i-\zi_j)
\prod_{i=1}^{r-1}\prod_{j,j'} (\zi_j -\zii_{j'})} \ .
\end{equation}
The polynomial $g(\bz)$ (which is symmetric under the exchange of
variables of the same color) vanishes when any of the following
conditions holds
\begin{align}
\zi_1 &= \zi_2 = \zii_1 \ , \qquad \zi_1=\zii_1=\zii_2 \ ,
\quad i=1,\ldots,r-1 \\
\zi_1 &= \zi_2 = \cdots = \zi_{l_i+1} = \zeta_i \ , \quad \forall i
\ .
\end{align}
Note that we do not impose the integrability conditions, because
$k$ will always be large enough, {\it i.e.\/} $k\geq \sum_i n^{(i)}$.
Finally, we need to impose the degree restriction on the function
$G(\bz)$. For each variable, we have that $\deg_{\zi_j}\leq 2$, which
follows form the fact that we use $\tfrac{1}{z}$ as a local variable
at infinity. In \cite{usII}, we showed that this leads to the
following polynomials
\begin{equation}
\label{kospol}
\cK_{\bl;\bn} (q) =
\sum_{\substack{\mi_a\geq 0\\i=1,\ldots,r\\
\bm=(\bC^{-1}_r) \cdot (\bn-\bl)}}
q^{\frac{1}{2} \mi_a (\bC_r)_{i,j} \bA_{a,b}\mj_{b}}
\prod_{i,a}
\begin{bmatrix}
\bA_{a,n^{(i)}} - (\bC_r)_{i,j} \bA_{a,b} \mj_{b} + \mi_a\\ \mi_a
\end{bmatrix}_q \ ,
\end{equation}
where $\bigl[ \; \bigr]_q$ denotes the $q$-binomial, which is defined to be
$\begin{bmatrix}n+m\\m\end{bmatrix}_q =
\frac{(q)_{n+m}}{(q)_{n}(q)_{m}}$ for $m,n\in \ZZ_{\geq 0}$ and zero
otherwise. The coefficient of $q^d$ in the expansion of
$\begin{bmatrix}n+m\\m\end{bmatrix}_q$ is the number of symmetric
polynomials of total degree $d$, in $m$ variables, where the degree of
each variable is maximally $n$. These $q$-binomials arise because of
the degree restrictions on the polynomials $G(\bz)$.
Note that the Kostka polynomials do not depend on the positions
$\zeta_i$, which follows from the fact that the degree of the rational
functions \eqref{kosratfunc} does not depend on these positions.

In \cite{usII}, we showed that the functions $\cK_{\bl;\bn}(q)$ of
\eqref{kospol} are polynomials in $q$, and are related in a simple way
to the generalized Kostka polynomials of \cite{SchWar,KirShi}.
In particular, by setting $q=1$, we obtain the Littlewood-Richardson
coefficients.

Let us make a few remarks about the structure of the polynomials
$\cK_{\bl;\bn}(q)$. Let $k(\bl) = \sum_i l_i$, which is the
lowest level at which $\lambda=\sum_i l_i \omega_i$ corresponds
to an integrable representation ({\it i.e.\/} $k(\bl)$ is the
threshold level). We then have the following results
\begin{equation}
\label{kosprop}
\begin{split}
\cK_{\bl;\bl} (q) & = 1\\
\cK_{\bl;\bn} (q) & = 0 \quad {\rm if} \quad
\begin{cases}
&k(\bl) > k(\bn)\\
&k(\bl)=k(\bn) \; {\rm and} \; \bl\neq\bn\\
&\sum_i i l_i \neq \sum_i i n^{(i)} \mmod r+1 \ .\\
\end{cases}
\end{split}
\end{equation}
These results are obtained by making use of the constraint
$\bm = (\bC_r)^{-1} (\bn-\bl)$ and the fact that all the summation
variables $\bm = (m_1,\ldots,m_r)$ have to be non-negative integers in
the sum in equation \eqref{kospol}.

We can view the polynomials $\cK_{\bl;\bn}$ as the entries of a
(square) matrix $\bK$, with entries $(\bK)_{\bl;\bn}(q) =
\cK_{\bl;\bn}(q)$. The relations \eqref{kosprop} imply that there is
an ordering such that the matrix $\bK$ is upper triangular with
$1$'s on the diagonal. Thus, the matrix $\bK$ is invertible.

Before we move on and give the character formul{\ae} for arbitrary
highest-weight representations, we will first give an example of the
`Kostka matrix' related to $\wsu(4)$. We use the following ordering of
the entries $\bl$
\begin{equation*}
(0,0,0);(1,0,1),(0,2,0);(2,1,0),(0,1,2);(4,0,0),(2,0,2),(1,2,1),
(0,4,0),(0,0,4)
\end{equation*}
With this ordering, we obtain the following Kostka matrix
\begin{equation}
\label{kexample}
\bK (q)=
\left(
\begin{array}{c|cc|cc|ccccc}
1& q&0& 0&0& 0&q^2&0&0&0\\
\hline
0& 1&0& q&q& 0&q&q^2&0&0\\
0& 0&1& 0&0& 0&0&q + q^2&0&0\\
\hline
0& 0&0& 1&0& 0&0&q&0&0\\
0& 0&0& 0&1& 0&0&q&0&0\\
\hline
0& 0&0& 0&0& 1&0&0&0&0\\
0& 0&0& 0&0& 0&1&0&0&0\\
0& 0&0& 0&0& 0&0&1&0&0\\
0& 0&0& 0&0& 0&0&0&1&0\\
0& 0&0& 0&0& 0&0&0&0&1\\
\end{array} \right)
\end{equation}
The inverse is
\begin{equation}
\label{kinvexample}
\bK^{-1}(q) =
\left(
\begin{array}{c|cc|cc|ccccc}
1& -q&0& q^2&q^2& 0&0&-q^3&0&0\\
\hline
0& 1&0& -q&-q& 0&-q&q^2&0&0\\
0& 0&1& 0&0& 0&0&-q - q^2&0&0\\
\hline
0& 0&0& 1&0& 0&0&-q&0&0\\
0& 0&0& 0&1& 0&0&-q&0&0\\
\hline
0& 0&0& 0&0& 1&0&0&0&0\\
0& 0&0& 0&0& 0&1&0&0&0\\
0& 0&0& 0&0& 0&0&1&0&0\\
0& 0&0& 0&0& 0&0&0&1&0\\
0& 0&0& 0&0& 0&0&0&0&1\\
\end{array}\right) \ .
\end{equation}

Using the results of the sections \ref{sec_kospol} and
\ref{sec_wzwcft}, we can now write
down explicit character formul{\ae} for arbitrary integrable
highest-weight representations of $\wsu (r+1)$, in terms of the
characters $\ch \cF_{\bn;k}^{\infty} (q,\bx)$ and the inverse matrix
$\bK^{-1} \bigl(\frac{1}{q}\bigr)$.

First of all, we have the result that we can decompose the characters
$\ch \cF^\infty_{\bn;k} (q,\bx)$ in terms of the characters of arbitrary
integrable highest-weight representations $\ch H_{\bl;k}(q,\bx)$ as follows
\begin{equation}
\ch \cF^\infty_{\bn;k} (q,\bx) = \sum_{\substack{\bl\\k(\bl)\leq k(\bn)}}
\cK_{\bl;\bn}\Bigl(\frac{1}{q}\Bigr) \ch H_{\bl;k} (q,\bx)\ .
\end{equation}
We can invert this relation, to obtain explicit character formul{\ae}
for arbitrary integrable highest-weight representations of $\wsu(r+1)$
\begin{equation}
\label{ahwrchar}
\ch H_{\bl;k} (q,\bx) = \sum_{\substack{\bn\\k(\bn)\leq k(\bl)}}
(\bK^{-1})_{\bn;\bl}\Bigl(\frac{1}{q}\Bigr) \ch \cF^\infty_{\bn;k} (q,\bx)\ .
\end{equation}
We would like to note that we have a similar formula for the character
of the principal subspace of general highest-weight representations
\begin{equation}
\label{ahwpsschar}
\ch W_{\bl;k} (q,\bx) = \sum_{\substack{\bn\\k(\bn)\leq k(\bl)}}
(\bK^{-1})_{\bn;\bl}\Bigl(\frac{1}{q}\Bigr) \ch \cF_{\bn;k} (q,\bx)\ ,
\end{equation}
where the character $\cF_{\bl;k}(q,\bx)$ is given by
\begin{equation}
\label{fchar}
\ch \cF_{\bl;k} (q,\bx) \stackrel{\rm def}{=}
(\prod_{i=1}^{r} x_i^{l_i})
\sum_{\substack{\mi_{a} \in \ZZ_{\geq0}\\a=1,\ldots,k\\i=1,\ldots,r}}
\Bigl(\prod_{i=1}^r x_i^{-\sum_j (\bC_r)_{ji}\mj}\Bigr)
\frac{q^{\frac{1}{2} \mi_a (\bC_r)_{i,i'} \bA_{a,a'} \mip_{a'} -
\bA_{a,l_i} \mi_a}}
{\prod_{i=1}^r \prod_{a=1}^{k} (q)_{\mi_a}} \ .
\end{equation}
This character can be viewed as the `untranslated' version of the
character \eqref{finfchar}, because applying the affine Weyl translation (as
explained in section \ref{sec_awt}), results in \eqref{finfchar}.

\def\cprime{$'$} \def\cprime{$'$} \def\cprime{$'$}

\vspace{.5cm}

\footnotesize
\sc{EA: Department of Physics, University of
    Illinois, 1110 W. Green St., Urbana, IL 61801.\\
\mbox{ardonne@uiuc.edu}}

\sc{RK: Department of 
    Mathematics, University of Illinois, 1409 W. Green Street, Urbana,
    IL 61801. rinat@uiuc.edu}

\sc{MS: Department of Physics, University 
      of Illinois, 1110 W. Green St., Urbana, IL 61801.\\
\mbox{m-stone5@uiuc.edu}}

\end{document}